%% file: ms.tex
\documentclass[12pt, preprint]{aastex}

\newcommand{\D}{\mathrm d}
\newcommand{\e}{\mathrm e}

\newcommand{\I}[1]{_{\mathrm #1}}

\newcommand{\Msun}{\mbox{$M_{\odot}$}}

\newcommand{\mc}{m_{\mathrm c}}
\newcommand{\mci}{m_{\mathrm c,0}}

\newcommand{\Dmc}{\Delta m_{\mathrm c}}
\newcommand{\me}{m_{\mathrm e}}
\newcommand{\beq}[1]{\vspace{-0.5mm} \begin{equation} \label{#1}}
\newcommand{\beqa}{\begin{eqnarray}}
\newcommand{\eeq}{\end{equation}}\vspace{-0.2mm}
\newcommand{\eeqa}{\end{eqnarray}}
\newcommand{\lsim}{\
\raise-2.truept\hbox{\rlap{\hbox{$\sim$}}\raise5.truept\hbox{$<$}\ }}
\newcommand{\gsim}{\
\raise-2.truept\hbox{\rlap{\hbox{$\sim$}}\raise5.truept\hbox{$>$}\ }}

\begin{document}

\title{New Optical and Near-Infrared Surface Brightness Fluctuations Models.
II. Young and Intermediate Age Stellar Populations}

\author{Raimondo G.\altaffilmark{1}, Brocato E.\altaffilmark{1},
Cantiello M.\altaffilmark{1,2}, Capaccioli M.\altaffilmark{3,4} }

\altaffiltext{1}{INAF--Osservatorio Astronomico di Teramo, Via M.
Maggini, I-64100 Teramo, Italy; raimondo@te.astro.it,
brocato@te.astro.it, cantiello@te.astro.it}


\altaffiltext{2} {Dipartimento di Fisica ``E.R. Caianiello'',
Universit\`a di Salerno and INFN, Sezione di Napoli, Gruppo
Collegato di Salerno, Via S. Allende, I-84081 Baronissi, Salerno,
Italy}

\altaffiltext{3}{Dipartimento di Scienze Fisiche, Universit\`a di
Napoli Federico II, Complesso Monte S. Angelo, via Cintia,
I-80126, Napoli, Italy; capaccioli@na.astro.it}

\altaffiltext{4}{INAF--Osservatorio Astronomico di Capodimonte,
via Moiariello 16, I-80131 Napoli, Italy}

\begin{abstract}

We present theoretical surface-brightness fluctuations (SBF)
amplitudes for single--burst stellar populations of young and
intermediate age ($25\, Myr \leq t \leq 5\, Gyr$), and
metallicities $Z=0.0003$, 0.001, 0.004, 0.008, 0.01, 0.02, and
$0.04$. The fluctuation magnitudes and colors as expected in the
Johnson--Cousins ({\it UBVRIJHK}) photometric system are provided.
We pay attention to the contribution of thermal--pulses asymptotic
giant branch (TP--AGB) stars. The sensitivity of the predicted SBF
to changes in the mass--loss scenario along the TP--AGB phase is
examined. Below $0.6$-$1\, Gyr$ both optical and NIR SBF models
exhibit a strong dependence on age and mass loss. We also evaluate
SBF amplitudes by using Monte Carlo techniques to reproduce the
random variation in the number of stars experiencing bright and
fast evolutionary phases (Red Giant Branch, AGB, TP--AGB). On this
ground, we provide constraints upon the faintest integrated flux
of real stellar populations required to derive reliable and
meaningful SBF measurements.

We analyze a technique for deriving SBF amplitudes of star
clusters from the photometry of individual stars, and estimate the
indetermination due to statistical effects, which may impinge on
the procedure. The first optical SBF measurements for 11 Large
Magellanic Cloud (LMC) star rich clusters - with age ranging from
few $Myr$ to several $Gyr$ - are derived by using Hubble Space
Telescope observations. The measurements are successfully compared
to our SBF predictions, thus providing a good agreement with
models of metallicity $Z = 0.0001$--0.01. Our results suggest
that, for TP--AGB stars, a mass loss as a power-law function of
the star luminosity is required in order to properly reproduce the
optical SBF data of the LMC clusters. Finally, near--infrared
models have been compared to available data, thus showing the
general trend is well fitted. We suggest how to overcome the
general problem of SBF models in reproducing the details of the
near--infrared SBF measurements of the Magellanic Cloud star
clusters.

\end{abstract}


\section{Introduction}

Surface brightness fluctuations (SBF) technique (Tonry \&
Schneider 1988, TS88) is one of the most powerful methods to
derive extragalactic distances of gas--free stellar systems. In
the last decade, SBF were proven to be effective in estimating
distances as far as 50 Mpc, and even larger, using ground--based
observations (e.g. Tonry et al. 2001; Blakeslee, Vazdekis, \&
Ajhar 2001; Mei et al. 2001; Liu Graham \& Charlot 2002), and out
to distances exceeding 100 Mpc from the space (Pahre et al. 1999;
Jensen et al. 2003). Since the method has been primarily applied
to elliptical galaxies and to the bulges of spirals, theoretical
SBF studies have been mostly oriented to old stellar systems ($t>
2$--$5 \, Gyr$: Worthey 1993; Buzzoni 1993; Liu, Charlot, \&
Graham 2000; Blakeslee, Vazdekis, \& Ajhar 2001; Cantiello et al.
2003).

Along with their ability in gauging distances, SBF appear to be a
very promising tool for investigating the evolution of unresolved
stellar populations in distant galaxies. Attempts have been made
to derive consistent estimations of age and metallicity for
samples of galaxies from SBF measurements (Brocato, Capaccioli \&
Condelli 1998; Liu, Charlot \& Graham 2000; Blakeslee, Vazdekis,
\& Ajhar 2001; Liu, Graham, \& Charlot 2002; Cantiello et al.
2003; Raimondo et al. 2004). However, even in galaxies dominated
by old stars, disregarding the effect of the contribution by
possible intermediate and young-age stellar populations ($t<
2$--$5 \, Gyr$) may be a hazardous assumption. The presence of
different stellar populations at different galacto-centric
distances, revealed by integrated colors and spectral--indexes
radial gradients, indicates that ellipticals are mostly composite
stellar systems (e.g. Pagel \& Edmunds 1981). SBF--gradient
measurements support this view (Sodemann \& Thomsen 1995;
Cantiello et al. 2005). The presence of relatively young
He--burning and/or AGB stars may contribute to the brightest part
of the galaxy luminosity function (LF). Because the major
contribution to the SBF signal comes from high luminosity stars,
it is relevant to know how SBF amplitudes change by lowering the
age of the stellar system down to a few million years, in those
photometric bands where the SBF signal is mostly affected by the
presence of young and intermediate-age stellar populations.

Before facing the problem of age and metallicity of possibly mixed
stellar populations in remote galaxies by using SBF, their
capability as population tracer has to be proven and carefully
tested on stellar systems of known distance, age, and metallicity.
Then, once calibrated on resolved Galactic and Local Group stellar
populations, SBF become a valuable tool in the analysis of the
stellar content in galaxies, where crowding and distance hamper
studies made with the classical color--magnitude diagram (CMD)
technique.

SBF have also been recognized effective in constraining stellar
evolution theory. In a previous paper \citep[hereafter Paper
I]{Cantiello+03}, we showed that SBF of old populations are
sensitive to the number of very bright stars evolving along the
early AGB, and thermally pulsating AGB (TP--AGB). In young stellar
populations --old enough to stars in these stages-- stochastic
fluctuations of the number of AGB--stars (as triggered by
mass--loss processes and evolutionary time-scale) is expected to
have even more relevance in predicting SBF amplitudes. If this is
confirmed, SBF can also provide a new and unexplored way for
improving our understanding of physical processes at work in AGB
stars of intermediate mass ($M\sim5\, M_\sun$).

In the last few years, a large effort to improve stellar evolution
models has been done in order to reproduce both the details of the
AGB--stars evolution \citep{Straniero+97, Pols+01}, and the
evolution of 'normal' stars. New physics experiments have advanced
our knowledge of nuclear reactions rates inside stars, and the
equation of state of stellar matter in critical conditions.
Updated and homogeneous evolutionary--track databases, reproducing
the observed CMD of young and intermediate--age stellar
populations in detail \citep[e.g.][]{Brocato+03}, are now
available \citep{Girardi+00, Castellani+03, Marigo+03,
Pietrinferni+04}. Therefore, very accurate SBF amplitudes can
be computed now in this age range.

Pioneer work on SBF from young Simple Stellar Populations (SSPs)
has been carried out by Gonzalez et al. (2003, 2004) in the
near-infrared (NIR), and by Raimondo et al. (2003) in the optical
regime. In the present paper, we evaluate SBF amplitudes expected
from SSPs younger than $5\, Gyr$, with metallicities from
$Z=0.0003$ to 0.04. Much attention is devoted to a simulation of
the TP--phase and its uncertainties, by evaluating changes of
chemical composition, stellar temperature, and the whole structure
of TP--AGB stars as prescribed by Wagenhuber \& Groenewegen (1998,
WG98). Mass--loss processes complicate enormously the modelling of
TP--AGB stars observational properties.

This complex picture is expected to have a huge impact on the SBF
behavior for intermediate--age stellar populations. In turn, SBF
magnitudes and colors could be decisive to put constraints on the
evolution of AGB stars, e.g. the efficiency of mass-loss, in
stellar systems with known age and metallicity, since they are
extremely efficient in mapping the properties of very bright stars
in the population (paper I).

In order to test our predictions, we select a sample of 11 star
clusters of the Large Magellanic Cloud (LMC): for 7 such clusters
the estimated age is within the range studied here; the others are
as old as the Galactic Globular Clusters (GGCs). Optical SBF
measurements are derived using the photometry of resolved stars
from high resolution images of each cluster, as obtained with the
Wide Field Planetary Camera 2 (WFPC2) on board of the Hubble Space
Telescope (HST).

The paper is organized as follows. A description of inputs of the
stellar population synthesis code is presented in
Section~\ref{section:Theoretical_Predictions}. The methods for
computing SBF amplitudes and stochastic effects, due to the number
of stars in the population are presented in Section
\ref{section:sbfamplitudes} and Section~\ref{section:statistics}.
SBF predictions are shown as a function of the TP--AGB stars
mass--loss rate in Section~\ref{section:sbfandtp}, and metallicity
in Section~\ref{section:sbfandmet}. We derive the optical SBF
measurements of the LMC star clusters in
Section~\ref{section:observations}, and present the relative
comparison with models in Section~\ref{section:sbf&LMC}, together
with a discussion on NIR SBF. A summary and final conclusions
end the paper (Section~\ref{section:conclusions}).

\section{Theoretical Framework}
\label{section:Theoretical_Predictions}

In order to compute SBF amplitudes, we use the  stellar population
synthesis code, described in Brocato et al. (1999, 2000). Here, we
only recall that the code starts directly from stellar
evolutionary tracks and relies on the Monte Carlo technique for
populating the Initial Mass Function (IMF). The former property
avoids any problem in the mass--bin or luminosity--bin procedures,
which may affect the use of isochrones. The latter property allows
us to take into account stochastic effects, due to the number of
stars in the stellar system, even for SBF amplitudes. If we deal
with star clusters and under--sampled stellar systems, stochastic
fluctuations of the number of bright stars may affect integrated
quantities \citep[e.g.][]{Santos&Frogel97, Brocato+99}. The
procedure used here takes directly into account these effects.

The present SSP models rely on the evolutionary--track database by
Pietrinferni et al. (2004, P04). All the evolutionary phases, from
the Main Sequence (MS) up to the AGB, are covered by models. In
particular, the AGB evolution runs up to the onset of the first
thermal pulse or to the carbon ignition. We selected stellar
evolutionary models with metallicities $Z=0.0003, 0.001, 0.004,
0.008, 0.01, 0.02$, and $0.04$, computed by adopting a
solar--scaled metal distribution with an enrichment law of $\Delta
Y/\Delta Z\simeq 1.4$.

In spite of the numerous improvements in the accuracy and
precision of stellar evolutionary models, only a few tracks
provide a detailed and full evolution along the TP--AGB phase
\citep[e.g. ][]{Straniero+97,Herwig+97,Pols+01}. Moreover, not all
of them cover homogeneously a wide range of chemical compositions
and stellar masses, needed to investigate resolved and unresolved
stellar populations. Since bright stars play a relevant role in
determining SBF, the TP--AGB phase cannot be neglected. We devoted
a particular care in including this evolutionary phase in our
simulations. For the sake of readability, the discussion and details on
how we threat TP--AGB stars are presented in Appendix A.

In the present paper, color--effective temperature ($T_{eff}$)
relations come from the semi--empirical compilation by Westera et
al. (2002), which is an updated version of the library by Lejeune
et al. (1997). A Scalo--like IMF \citep{Scalo98} is assumed for
stellar masses in the interval $m=0.1$--$10\, M_{\sun}$. Note that
the upper mass limit corresponds to the highest mass evolving off
the MS in the youngest population we considered, i.e. $t\sim 25 \,
Myr$. All the more massive stars are expected to be exploded as
supernovae, since their quiescent nuclear burning life-time is as long
as a few million years.

\begin{figure*}
\epsscale{1} \plotone{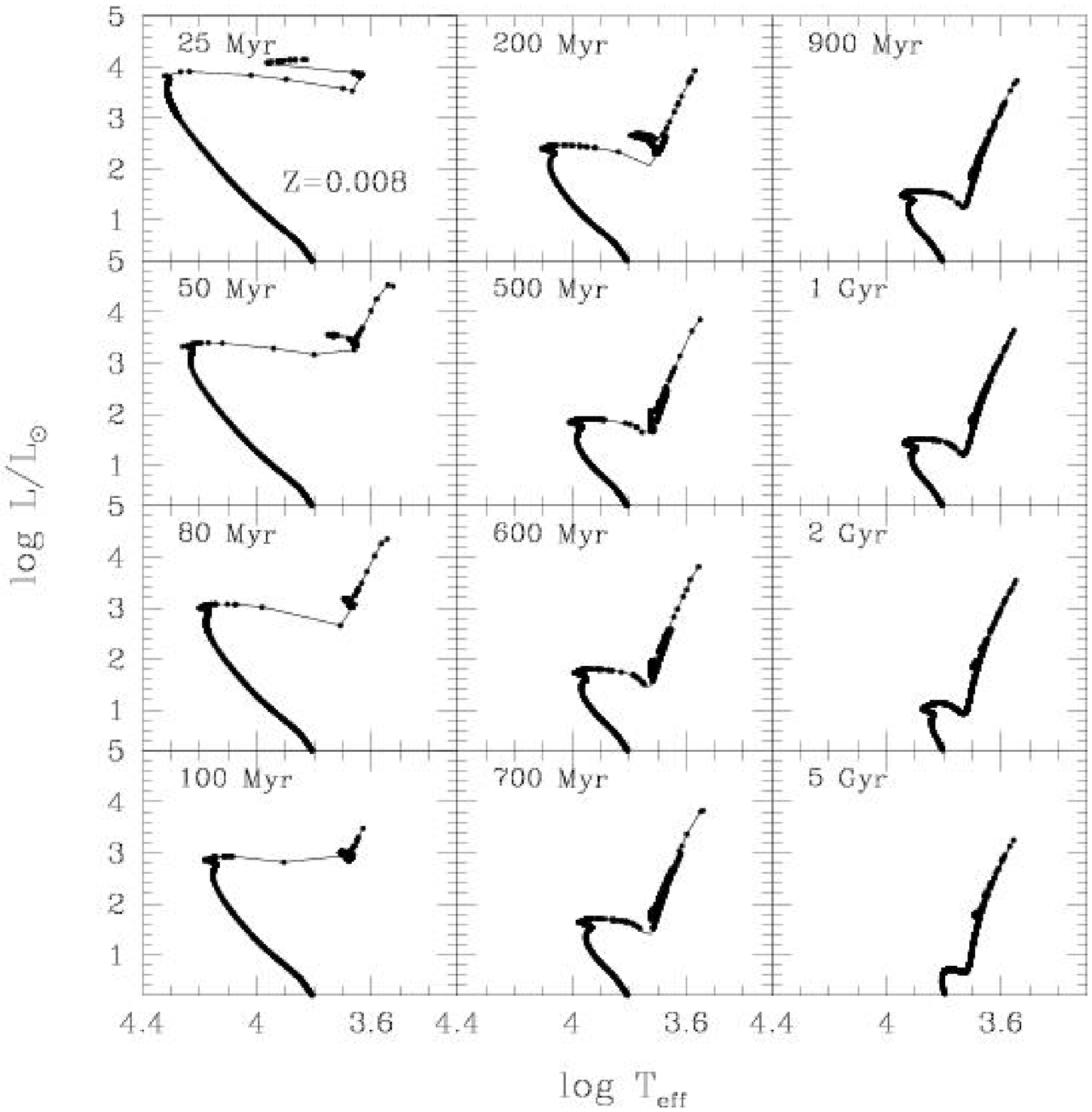} \caption{A sample of synthetic $log\
L/L_{\sun}$ $vs.$ $log\ T_{eff}$ diagrams for the labelled ages
and $Z=0.008$. In each panel one of the 5000 synthetic CMDs used
to compute the SBF is plotted (dots); the line represents the
isochrone. $B1$ mass--loss rate is assumed (see text and Appendix
A).} \label{fig:cmd}
\end{figure*}

By means of Monte Carlo techniques, $N_{sim}=5000$
\emph{independent} simulations are computed for each set of SSP
parameters, i.e. age ($t$), and metallicity ($Z$). The total mass
of each simulation is typically $ {\cal {M}} \simeq 10^4 \,
M_{\sun}$, unless explicitly stated otherwise. The explored age
range is $25\, Myr \leq t \leq 5\, Gyr$\footnote{At the time of
publication, models fully consistent with the present theoretical
scenario for ages larger than $5 \, Gyr$, and for all
metallicities presented here are available at the web site:
www.te.astro.it/SPoT. These old--age models will be discussed in
detail in a forthcoming paper.}. As an example of one of the 5000
simulations, in Fig.~\ref{fig:cmd} we report the synthetic CMDs of
the simulated stellar populations -- for few selected ages -- in
the theoretical $log \ L/L_{\sun}$ $vs.$ $log \ T_{eff}$ plane.

\section{SBF Amplitudes}
\label{section:sbfamplitudes}

In this section, we present two different methods for computing SBF.
We discuss in detail their analogies with observations, and
investigate stochastic effects on SBF amplitudes, caused by the
discrete nature of the number of stars in stellar systems. The
impact of bright and rare TP--AGB stars on SBF is also presented.

The \emph{standard} procedure we developed to calculate SBF is
already presented in Paper I ($std$--procedure). It is based on
the following equation, valid in the \emph{gaussian statistics}
regime, i.e. high number of stars (TS88):

\begin{equation}
\overline {M}_{X} = -2.5 \, log \, \overline {F}_{X} = -2.5 \, log
\, [\frac{ \langle { (\, F_{X} - \langle {F_{X}} \rangle
)^2}\,\rangle }{\langle F_{X} \rangle} ] \label{eq:ourformula}
\end{equation}
where $\overline {F}_{X}$ is the fluctuation flux in the generic
photometric filter $X$; ${F}_{X}$ and $\langle {F}_{X} \rangle$
are respectively the total flux of each simulation, and the mean
total flux averaged over $N_{sim}$ simulations, i.e.:

\begin{equation}
{F}_{X} \equiv {F}_{X}^j ={\sum_{i=1}^{N_{star}} f_i(X)}
\;\;\;\;\;\;\; j=1, N_{sim} \label{eq:ourformula2}
\end{equation}
$f_i(X)$ corresponds to the absolute flux of the $i$-th star
populating the $j$-th simulation, and

\begin{equation}
\langle F_{X} \rangle
 =\frac{{\sum_{j=1}^{N_{sim}}} {F}^{j}_{X}}{N_{sim}}
\label{eq:ourformula3}
\end{equation}

For the sake of clarity, in this section absolute SBF amplitudes
derived from Eq.~\ref{eq:ourformula} are called
\emph{standard}--SBF, and indicated as $\overline {M}_X^{std}$. As
quoted in Paper I, there is a close correspondence between the
\emph{std}--procedure and the way of measuring SBF for unresolved
stellar populations. In fact, the integrated energy flux
${F}_{X}^j$ corresponds to the flux measured in a single pixel of
a galaxy CCD image (if both seeing and population mixture are
neglected).

When $N_{sim}$ is equal to 5000, $\overline {M}_X^{std}$ is called
\emph{asymptotic value} ($\overline {M}_X^{std,asym}$). Note that
$\overline {M}_X^{std}$ runs very quickly (after few hundreds of
simulations) towards $\overline {M}_X^{std,asym}$ (Appendix B),
and the uncertainties are reduced ($\simeq 0.05\, mag$,
Paper I). The \emph{asymptotic} value has the same physical
meaning of the classical SBF predictions in the
literature (e.g. Worthey 1993; Buzzoni 1993; Liu, Charlot, \&
Graham 2000; Blakeslee, Vazdekis, \& Ajhar 2001, G04).

For spatially-resolved star clusters, we can apply another technique for
measuring SBF. It was first introduced by Ajhar
\& Tonry (1994, AT94), and is based on individual--star
photometry. By means of our method of computing SSP models and
integrated properties, we can provide SBF predictions for
each $j$-th simulation, by directly applying the definition of SBF,
as introduced by TS88 (Eq. 7-9):

\begin{equation}
\overline {M}^{RS,j}_X = -2.5 \log \ [ \frac
{\sum_{i=1}^{N_{star}} f_i(X) ^2} {\sum_{i=1}^{N_{star}} f_i(X)} ]
\;\;\;\;\;\;\; j=1, N_{sim}. \label{eq:eqTS}
\end{equation}
Note that $\overline {M}^{RS,j}_X $ corresponds to the SBF
obtained using the photometry of all stars in a cluster, and
directly relies upon the \emph{poissonian statistics}. The mean SBF
magnitude averaged over $N_{sim}=5000$ independent simulations is:

\begin{equation}
\overline {M}^{RS}_X =   \frac {\sum_{j=1}^{N_{sim}} \overline
{M}_X^{RS,j} } {N_{sim}}, \label{eq:eqTS2}
\end{equation}
and the SBF statistical uncertainties can be derived as the
standard deviation of the $\overline {M}_X^{RS,j}$ distribution.
In the following, this procedure for computing SBF is indicated as
\emph{RS}--procedure (Resolved Systems).

\subsection{Statistical Effects} \label{section:statistics}

In principle, the two procedures should provide very similar
results. In practice, this is only true if the number of stars
included in the $j$-th simulation is large enough to populate all
the evolutionary phases (i.e. the poisson statistics coincides
with the gaussian one, TS88; Paper I). Hence, in order to analyze star
clusters SBF magnitudes, and to properly compare models with
observations, it is crucial to investigate the dependence of  SBF
amplitudes upon quantities related to the number of stars in the
population. In the following, we present a careful analysis of
the statistical effects on SBF predictions. Although the analysis is
specifically performed for star clusters, the technique we
developed can be applied to any spatially resolved populations.

For fixed age and metallicity, variations of the total number of
stars among clusters result in a variation of the cluster total
$V$--magnitude ($M_V^{tot}$). This is an observational quantity
related to the number of stars included in the stellar population.
We computed SBF amplitudes, by adopting both procedures
(\emph{std}--procedure: Eq.~\ref{eq:ourformula}; and
$RS$--procedure: Eq.~\ref{eq:eqTS2}), by varying the SSP
integrated absolute magnitude ($M_V^{Tot}$) from
$M_V^{Tot}\simeq-1$ to $-11\, mag$, a range larger than that
observed for LMC clusters. This corresponds to varying the total mass
of the population from ${\cal M}_{tot}\sim 10^2$ to $\sim 10^5
M_{\sun}$. As usual, for each set of parameters ($t$, $Z$) we
computed $N_{sim}$ \emph{independent} simulations.

\begin{figure*}
\plotone{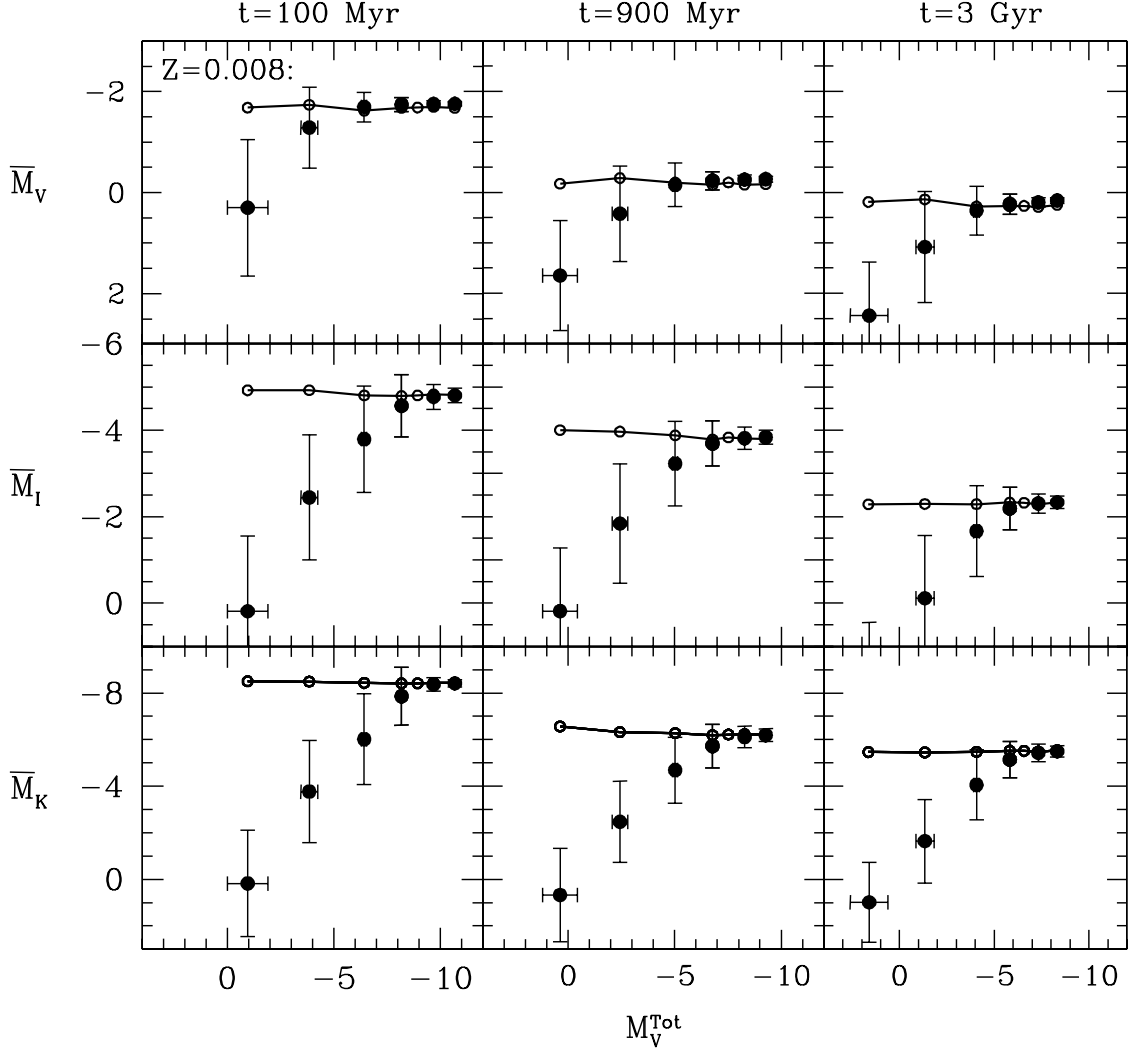} \caption{SBF absolute magnitudes as a function of
the integrated absolute magnitude $M_V^{Tot}$. The thick solid
lines (open circles) represent the \emph{asymptotic SBF}
($\overline {M}^{std,asym}_X$, Eq.~\ref{eq:ourformula}). Filled
circles represent $\overline {M}^{RS}_X$ obtained from
Eq.~\ref{eq:eqTS2} (see text). Adopted metallicity and ages are
labelled. } \label{fig:statistics1}
\end{figure*}

Fig.~\ref{fig:statistics1} shows the behavior of $\overline
{M}^{std,asym}_X$ in the $V$, $I$, and $K$ bands as a function of
$M_V^{Tot}$ (solid thick line, open circles). Models for three
different ages $t=100\, Myr$, $t=900\, Myr$, and $t = 3\, Gyr$,
and $Z=0.008$ are displayed. From all the panels, it is evident
that \emph{asymptotic} SBF amplitudes ($\overline
{M}_X^{std,asym}$) are stable within $0.1 \, mag$ in the
wide range of $M_V^{Tot}$. In other words, independently of
the cluster mass (luminosity), fluctuation amplitudes computed
over 5000 simulations with the \emph{std}--procedure predict
nearly constant values.

SBF amplitudes resulting from the \emph{RS}--procedure ($\overline
{M}_X^{RS}$) are plotted as filled circles in
Fig.~\ref{fig:statistics1}. The uncertainties due to stochastic
fluctuations of the number of bright stars are directly evaluated
as the standard deviation of the $\overline {M}^{RS}_X$
distribution, and plotted as 1 $\sigma$ error-bars. It is worth
noting that the $\overline {M}^{RS}_X$ \emph{versus} $M_V^{Tot}$
behavior is similar to that shown by integrated colors
\citep[e.g.\ ][]{Santos&Frogel97,Brocato+99}.

The difference between the SBF \emph{asymptotic} value and that
derived from Eq.~\ref{eq:eqTS2} is larger in NIR bands than in
blue bands, because in the optical bands the contribution of the
sparsely populated RGB and AGB is less important than that of the
well populated MS. By increasing the total luminosity (i.e.
increasing SSP mass), the two procedures converge to the same SBF
value, and the uncertainty due to stochastic effects decreases
accordingly. This happens when each simulation is well populated
in all post--MS evolutionary stages until the TP--phase. The
maximum value of $M_V^{Tot}$ ($M_V^{Tot,max}$), where the two
procedures give the same results --within uncertainties-- slightly
depends on the age, in the sense that the older the population,
the fainter is $M_V^{Tot,max}$. It changes from
$M_V^{Tot,max}\simeq -6.5$ to $-4.5\, mag$ by varying the age from
$t=100\, Myr$ to $3\, Gyr$.

Since the two theoretical procedures reflect different ways of
measuring SBF, the last finding has direct effects upon the
observations. The condition $M_V^{Tot} \lsim M_V^{Tot,max}$ must
be satisfied when comparing classical (\emph{asymptotic}) SBF
predictions to measurements derived from single--star photometry.
In relatively poorly populated stellar systems, the random
occurrence of bright stars deeply affects the SBF, and should be
taken into account in the comparison with models (that is
$\overline {M}^{RS}_X$ should be preferred).

In the \emph{std}--procedure (i.e. the one applied to measure SBF in
galaxies) SBF are derived by using the integrated flux of a large
number of 'similar' but 'uncorrelated' populations (i.e. 'similar'
but 'uncorrelated' pixels). This ensures the \emph{asymptotic} SBF
value is reached even taking into account a relatively small
number of pixels  (see Appendix B). For example, if a fraction of
young populations is expected in a galaxy,
simulations suggest that the effective number of pixels used in
measuring SBF should not be less than 2000, so as to reduce the
uncertainties due to statistical effects. This constraint becomes
less stringent if the mass density of the young population is
larger than $\gsim 10^4\, M_{\sun}/ $pixel, i.e. 500 pixels are
expected to be enough to keep such uncertainties below $0.1\,
mag$.

In conclusion, we find that SBF models (classical asymptotic
models) well represent SBF as observed in galaxies, while SBF
measured from resolved stellar systems require caution before
performing comparisons with models. In particular, when comparing
SBF predictions with SBF observations for star clusters or
under--sampled stellar populations, attention should be payed in
taking into account uncertainties due to the stochastic occurrence
of bright stars experiencing fast and luminous evolutionary
phases.

\subsection{The impact of TP--AGB stars}
\label{section:sbfandtp}

Being extremely bright and rare, TP--AGB stars are relevant in
determining SBF of young/intermediate--age stellar populations.
One of the processes which triggers the luminosity and the
duration of the TP-phase is mass--loss. There are different
mass--loss scenarios to be adopted along the TP--phase (see
Appendix A for a detailed discussion). This is a critical
assumption, because the mass--loss efficiency determines the
number of very bright TP--AGB stars. In order to understand the
effective (quantitative) impact of such stars on SBF, we computed
a set of models by varying only the mass--loss rate:

$BH$--models: \emph{mild} mass--loss rate (Baud \& Habing 1983,
hereafter BH), the TP--phase is well populated (Eq.~\ref{eq:bh}) ;

$B1$--models: \emph{moderate} mass--loss rate (Blocker 1995,
hereafter B95), few stars in the TP--phase (Eq.~\ref{eq:b1});

$B2$--models: \emph{high} mass--loss rate (B95), few stars in the
TP--phase (Eq.~\ref{eq:b2});

$no$--$TP$ models: \emph{huge} (unreal) mass--loss rate, no stars
in the TP--phase.

\begin{figure*}
\epsscale{1} \plotone{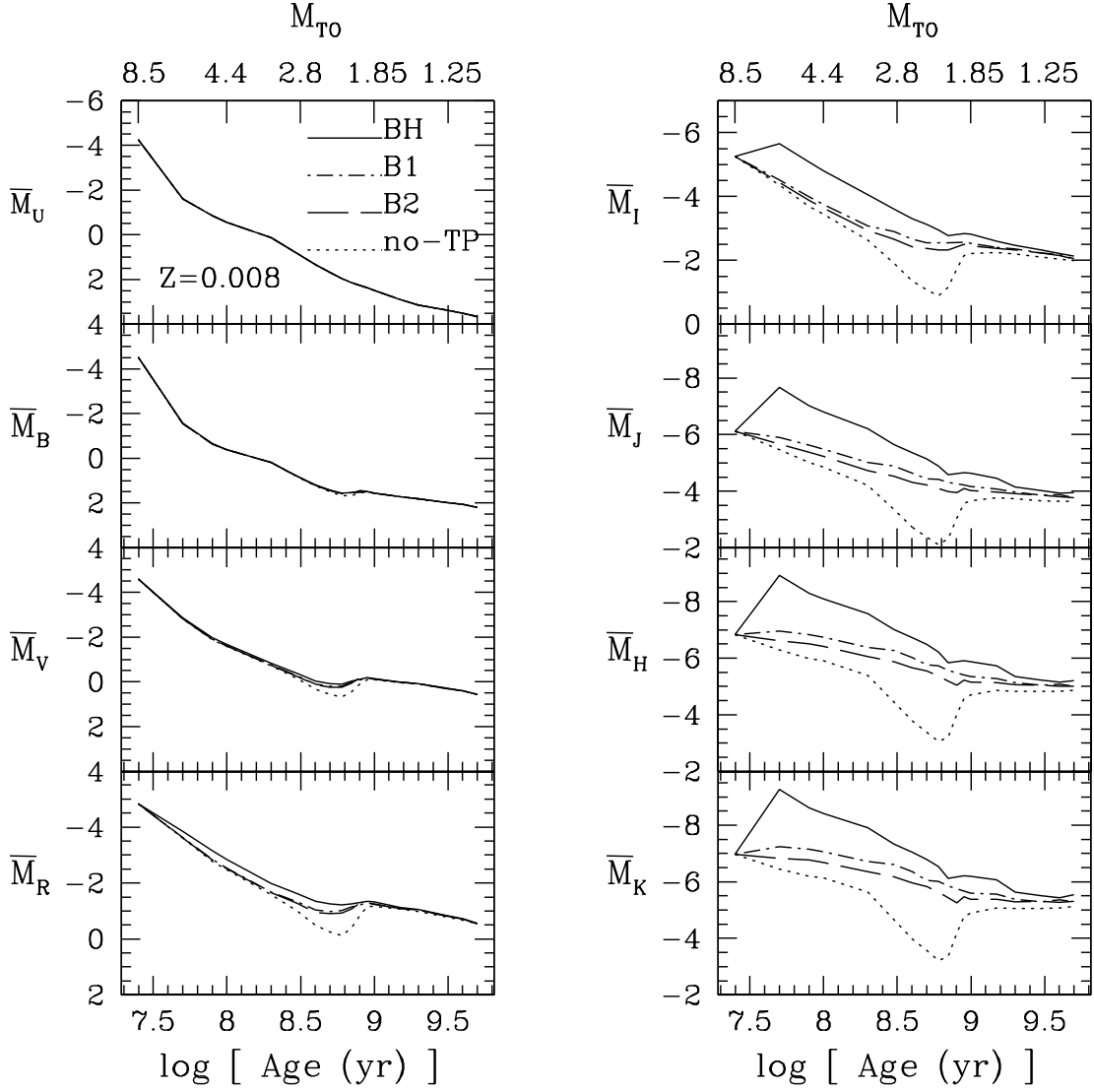} \caption{SBF magnitudes as a
function of age, and fixed metallicity ($Z=0.008$). Dotted lines
correspond to models computed neglecting the TP--AGB phase
($no$--$TP$). Models including TP stars are plotted for three
different mass--loss rate assumptions: $BH$--models (solid lines),
$B1$--models (dot--dashed lines), and $B2$--models (long--dashed
lines). The MS turn--off masses at selected ages are also
labelled.} \label{fig:sbfvsage}
\end{figure*}

Let us analyze the case of $Z=0.008$, i.e. a metallicity
expected to be representative of young/intermediate LMC star
clusters \citep{Westerlund97} we shall use as observational
counterparts in Section~\ref{section:observations}.
Fig.~\ref{fig:sbfvsage} illustrates the time evolution of SBF
predictions for the four assumptions of mass--loss rate. We find
that SBF in the $U$, $B$, and $V$ bands are not significantly
affected by TP--AGB stars mass--loss processes. Whereas, for
redder bands, from $R$ to $K$, SBF may change by more than $1\,
mag$ for models older than $\sim 50\, Myr$ and younger than $\sim
1\, Gyr$. The former value corresponds to the appearance of the
AGB in an SSP of such a metallicity (AGB--phase transition); the
last one is related to the appearance of the RGB \citep[RGB--phase
transition, ][]{ Renzini&Buzzoni86}. In this age--range, the cool
AGB (including TP--AGB) stars dominate the integrated bolometric
light.

In $no$--$TP$ models, the brightest objects in the SSP are
early--AGB stars, i.e. stars before the first thermal pulse. Thus,
SBF amplitudes are less luminous and extremely sensitive to the
appearance of the RGB. On the other hand, $BH$ mass--loss rate is
not very efficient, and the resulting relatively high number of
TP--stars has three consequences: i) SBF are the most luminous
(within our four cases) at any age, ii) SBF experience an evident
jump, due to the occurrence of the AGB phase--transition at $t \sim
50\, Myr$, and iii) SBF do not vary significantly at the RGB--phase
transition age. $B1$-- and $B2$--models lie within the previous
two extreme cases. Since $B1$ and $B2$ mass--loss rates
exponentially depend upon the star luminosity, mass loss is much
more efficient in massive AGB stars (young SSPs, $t\lsim 200\,
Myr$) than in low--mass AGB stars (old SSPs). Consequently, in
young SSPs, the TP--phase is less populated, thus bringing the SBF values
toward the $no$--$TP$ case.

For ages $t\gsim 1\, Gyr$ AGB stars become less relevant in
predicting SBF, because the RGB tip is nearly as bright as the AGB
tip, but much more populated (of about a factor 10). This is the
main reason why the four curves appear to converge at older ages.

\begin{figure*}
\epsscale{1} \plotone{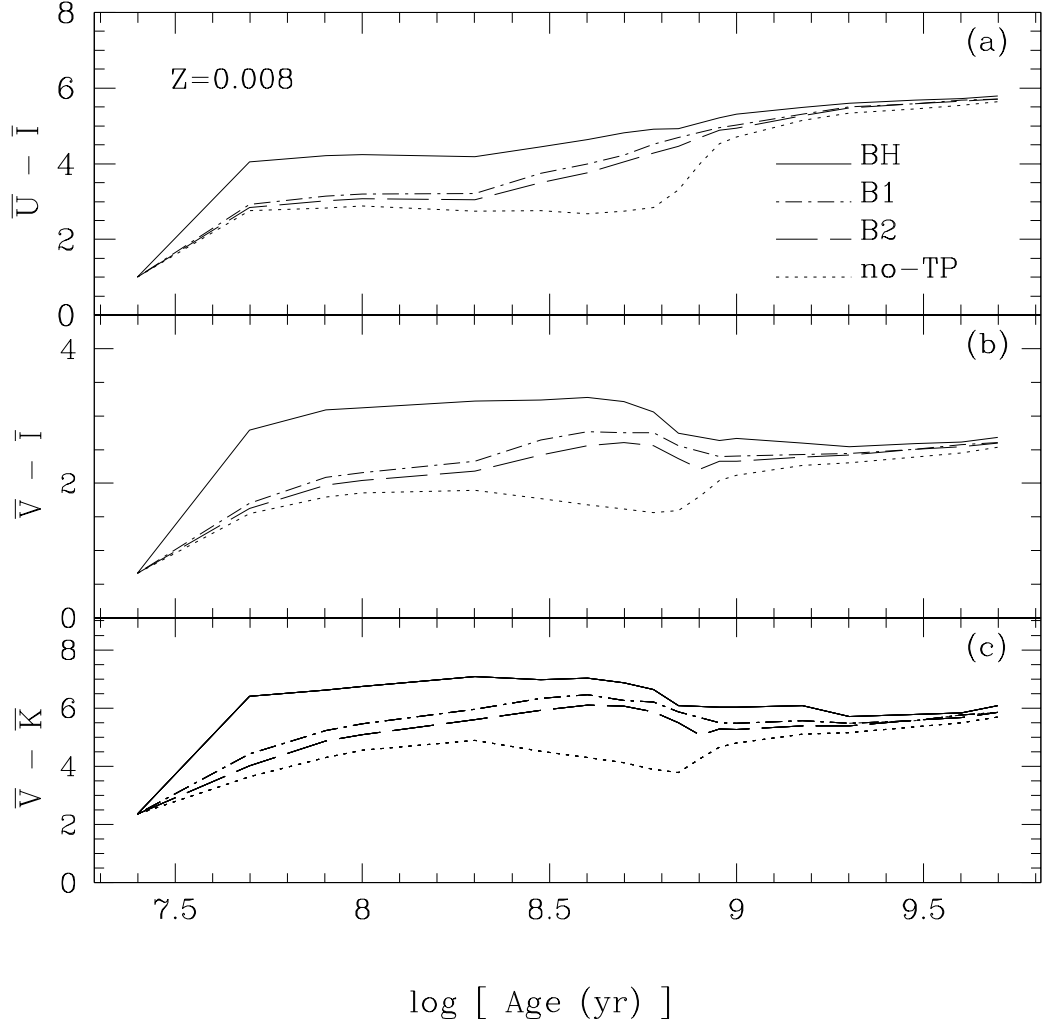} \caption{$\overline {U}-\overline
{I}$, $\overline {V}-\overline {I}$, and $\overline {V}-\overline
{K}$ fluctuation colors as a function of age, and fixed
metallicity ($Z=0.008$). Symbols are as in
Fig.~\ref{fig:sbfvsage}. The sensitivity to mass--loss rate
assumed in modelling the TP--AGB phase is evident in the plotted
colors. } \label{fig:col1}
\end{figure*}

Fluctuation colors present a very similar behavior
(Fig.~\ref{fig:col1}), showing a high sensitivity to TP--AGB stars
mass--loss rate and to the phase--transitions. Of course, SBF
colors might be more effective for detecting the efficiency of
TP--AGB stars mass--loss rate than fluctuation magnitudes, as a
consequence of their independence from distance.

The results in Fig.~\ref{fig:sbfvsage} and Fig.~\ref{fig:col1}
suggest at least two considerations: $a)$ indications on the age
of distant stellar systems can be inferred by measuring SBF in
blue bands ($U$, $B$ and $V$); and $b)$ SBF measured in
single--burst stellar populations would provide a tool to
quantitatively evaluate the properties of TP--AGB stars, e.g. the
expected number, luminosity, and their mass loss.

\subsection{SBF versus Metal Content} \label{section:sbfandmet}

\begin{figure*}
\epsscale{1} \plotone{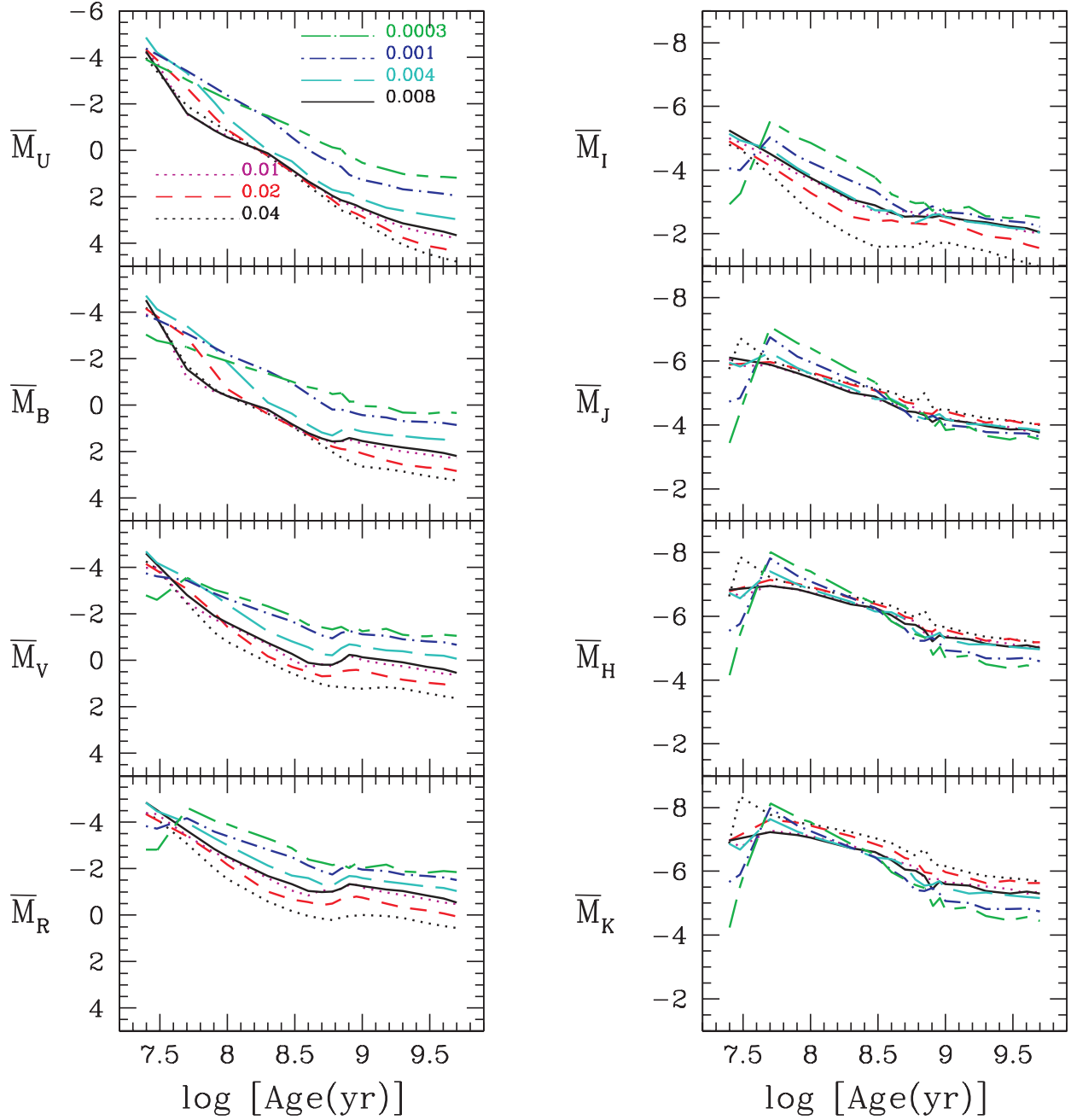} \caption{Time evolution of SBF
magnitudes for different metallicities ($B1$--models). The
different lines represent the labelled metallicity values.}
\label{fig:sbf_met}
\end{figure*}

Before concluding this Section, we discuss the SBF dependence on
the chemical composition. Table~\ref{table:table_sbf} and
Fig.~\ref{fig:sbf_met} report SBF amplitudes in Johnson--Cousins
bands for different chemical compositions as a function of age.
The input assumptions are the same as quoted at the beginning of
this section, and the adopted mass--loss scenario is $B1$. The
reason of this choice is given in
Section~\ref{section:observations}. Table~\ref{table:table_sbf}
lists: age (Col. 1), absolute SBF magnitudes in various
photometric filters (Cols. 2--10), absolute integrated $V$
magnitude and $V-I$ integrated color of the population (Cols.
11--12). The SBF predictions are available at the web--site:
http://www.te.astro.it/SPoT.

The general trend of SBF magnitudes at different chemical
compositions is similar to the one shown by models with $Z=0.008$.
As a general indication, we find that metal--poor ($Z\lsim 0.001$)
SSPs tend to have brighter SBF in the optical bands. It is the
opposite for NIR bands, since metal--rich populations show
brighter SBF, especially at an old age. This can better understood
if we remember that the RGB and AGB of metal--rich stars are
cooler than the corresponding branches of metal--poor stars.
Moreover, at young ages (several Myr) the appearance of AGB stars
causes a sudden jump in the \emph{mean} NIR--SBF magnitude/color
of the population. This is because SBF are an extremely efficient
measure of any fluctuation of the distribution
 of bright stars in the population. Since  metal--rich
(pre-AGB) bright stars are typically \emph{redder} than similar
metal--poor stars, correspondingly, the appearance of AGB stars is
expected to produce a less intense variation in the NIR SBF of a
metal--rich population.

Looking at both Fig.~\ref{fig:sbfvsage} and
Fig.~\ref{fig:sbf_met}, one can note that the strong dependence of
NIR SBF on mass loss, especially for intermediate--age
populations, has several implications. The NIR SBF with
intermediate mass--loss (B1) can look similar to those with mild
mass--loss (BH), depending on metallicity. This behavior might
generate a possible problem of degeneracy, when the metallicity of
the stellar system is unknown. In addition, the evidence that mass
loss increases with metallicity in AGB stars
\citep{Groenewegen+95} might complicate the scenario. On the other
hand, this high sensitivity can be used to discriminate among
different mass--loss assumptions, if stellar systems with known
metallicity are considered, thus making SBF an interesting tool
for investigating properties of AGB stars. In
Section~\ref{section:observations}, we shall use WFPC2/HST SBF
data to discriminate the mass--loss scenario, active in TP--AGB
stars, for a sample of LMC star clusters.

\section{SBF Measurements}
\label{section:observations}

To our knowledge, optical--SBF models available in the literature
do not include predictions for young stellar populations, as they
usually extend down to 2--$5\, Gyr$ (Worthey 1993; Buzzoni 1993;
Liu, Charlot, \& Graham 2000; Blakeslee, Vazdekis, \& Ajhar 2001;
Paper I). Thus, no comparison with optical SBF--models from other
authors can be done. In this Section, we provide the first step
toward the measure of optical SBF for young simple stellar
systems.

Stellar clusters in the Magellanic Clouds (MC) represent a
unique opportunity to explore the behavior of SBF in young and
intermediate--age stellar systems. The MC clusters --formed at
different epochs-- provide a remarkable sampling of stellar
clusters in a wide range of ages. Moreover, because of their
proximity, they allow us to perform direct photometry of individual
stars. This is of paramount importance in probing stellar
population synthesis tools and models. The criteria for selecting
MC star clusters, as well as photometric data analysis, are two important
steps in our overall investigation. Hence, in the following
section, we describe these steps in detail before discussing the
SBF measurements.

\subsection{Star clusters and photometric data selection}
\label{section:clusters_selection}

As shown in Section \ref{section:statistics}, the low statistic in
the brighter part of the cluster luminosity function (LF) plays a
role in determining the uncertainties of the measured SBF. For
this reason, we prefer clusters with a number of stars high enough
to avoid large statistical fluctuations
(Fig.~\ref{fig:statistics1}). While taking this condition into
account, we give the priority to more massive clusters, by
selecting LMC young clusters, whose integrated magnitudes satisfy
the condition $M_V^{Tot} \lsim M_V^{Tot,max}$ at any age.

An accurate photometry of individual stars in the cluster core is
another crucial requirement in order to fully map all the features
of the cluster stellar population. We take advantage of the
WFPC2/HST high capability of resolving stars in the core of the MC
clusters. Mackey \& Gilmore (2003, MG03) have shown that the core
radius of MC clusters is typically smaller than 30\arcsec, thus
the WFPC2 field of view is large enough to contain most stars in
the cluster.

With a total apparent magnitude $V^{Tot}=9.89\pm 0.01$
\citep{vandenBergh81}, NGC 1866 is one of the most massive cluster
formed in the LMC during the last $3\, Gyr$. Our group has
recently obtained deep and accurate WFPC2/HST observations in the
$F555W$ ($\sim V$) and $F814W$ ($\sim I$) filters of this cluster
\citep{Walker+01,Brocato+03}. Hence, it is a good candidate for
our purpose. Further, we select a sample of LMC clusters spanning
the age range from a few million up to a few billion years, for
which similar HST observations in the same photometric filters are
available. A sub--sample of the LMC clusters studied by de Grijs
et al. (2002a, 2002b, 2002c) satisfy most requirements, namely:
NGC~1805, NGC~1818, NGC~1831, NGC~1868, NGC~2209 and $Hodge\ 14$.
Their observations, just like our own for NGC 1866, reach
magnitudes as faint as $V\sim 25\, mag$. Moreover, de Grijs and
collaborators could detect radial mass segregation in the central
regions of these LMC clusters. This is a clear indication of the
high level of completeness and accuracy of their photometric data
and LFs.

In order to minimize possible differences in handling the data, the
original images have been retrieved from the HST archive, and
analyzed by following the same procedure - discussed in Brocato et al.
(2003), and briefly described here. The basic information on the
images we used are summarized in Table \ref{tab:hstimages}. Each
frame has been pre--processed according to the standard WFPC2
pipeline, by using the latest available calibrations. The
removal of cosmic rays and the photometry have been performed
by using the most recent version of the HSTphot package developed by
Dolphin (2000a). The Point Spread Function (PSF) fitting option on
the HSTphot routine has been adopted in order to take advantage of the PSFs,
which  are computed directly to reproduce the shape details of
star images as obtained in the different regions of the WFPC2.
Charge Transfer Efficiency (CTE) corrections and calibrations to
the standard $VI$ system were obtained directly by HSTphot
routines, as documented by Dolphin (2000b). In
Fig.~\ref{fig:cmd_clusters} the resulting CMDs are plotted,
together with the typical uncertainties on the photometry as a
function of magnitude. The $V$ and $I$ photometry of de Grijs and
collaborators are compared with the present work. The agreement is
extremely good, being the mean differences of the order of a few
hundreds of magnitudes in all the chips, and for all the clusters.
Completeness has been evaluated by distributing artificial stars
of known positions and magnitudes, in selected circular regions
around the cluster center. Since the resulting completeness
functions strictly agree with those published by de Grijs et al.
(2002a, 2002c, see their Fig. 2), we do not present nearly
identical figures.

Even if the present paper is mainly devoted to young stellar
clusters, for the sake of completeness we also added four really
old clusters. We derived $\overline V$ and $\overline I$ also for
NGC~1754, NGC~1916, NGC~2005, and NGC~2019, by relying on the high
quality and deep HST photometry by Olsen et al. (1998), who used
deep exposures in both $F555W$ and $F814W$ filters. This assures
to measure stars of $V \sim 25$, well below the MS TO point for
all the clusters.

Finally, we have a sample of 11 LMC star clusters. The complete
list and a few properties of the clusters are presented in
Table~\ref{table:table_lmc}: cluster name (Col.\ 1); total
$V$--magnitude from van den Bergh (1981), and present work (Cols.\
2, 3); $\overline V$, $\overline I$, and $\overline V - \overline
I$ (Cols.\ 4-6); Col.\ 7 lists cluster age as derived in
Section~\ref{section:Age_determination} or from the literature
(references in Col.\ 8). The uncertainties of SBF data refer to
the maximum/minimum SBF values obtained by including field
contamination and 'missed' bright stars, as described in the
following section. The integrated magnitudes obtained in the
present work refer to the photometric data used to derive the SBF
measurements. They agree with values of van den Bergh (1981)
within few tenths of magnitudes. For $Hodge\, 14$ the difference
is larger, due to the severe area selection we used in order to
minimize the field contamination (see the following section).

\subsection{Optical SBF}
\label{section:optical}

In order to measure SBF, we followed the procedure described by AT94,
by means of the photometry of individual stars in the cluster.
Stars 8-10 $mag$ fainter than the brightest stars of the cluster
have been measured through high--resolution WFPC2 imaging. This
allows us to use individual stars photometry not only to evaluate
the numerator in Eq. \ref{eq:eqTS}, but also the denominator
without introducing other sources of uncertainty in estimating the
total flux (e.g. sky--level evaluation). Differently from the
present work, AT94 (and G04) were forced "...to sum the flux of
the CCD image with the sky subtracted..." to account for the
contribution of the large number of unmeasured faint stars to the
total flux. This is due to the fact that the photometry used in those papers
does not reach the faint part of the MS (see e.g. the case of 47
Tuc in Fig. 10 of AT94).

The faint magnitude limit of the photometry is a crucial point in
evaluating the denominator of Eq. \ref{eq:eqTS}, as enlightened by
AT94. In order to find a quantitative indication of the impact of
faint MS stars on SBF, we used the following procedure. First, the
$VI$ SBF are derived from synthetic $VI$ CMDs, containing stars
with masses down to $M=0.1\, M_\sun$, by applying the
$RS$--procedure (Eq.~\ref{eq:eqTS}). Then, the SBF are re-computed
by artificially cutting out stars with $V > V_{cut}$ from the
complete synthetic CMDs for a selected set of $V_{cut}$. This test
corresponds to a simulation of $100\%$ incompleteness at
magnitudes fainter than $V_{cut}$. The $V$ and $I$ bands SBF have
been derived by adopting the LMC absolute distance modulus
$(m-M)_0 =18.4 \pm 0.1$, and mean reddening $E_{B-V}=0.06$
\citep{Walker+01, Alcock+04}. The procedure has been repeated for
three different populations aged: $t = 100\, Myr$, $900\, Myr$,
and $4\, Gyr$. The differences between the SBF derived by
considering the complete synthetic CMDs and those from the
'cutted' ones are presented in Fig. \ref{fig:sbfVcut} as a
function of $V_{cut}$. The figure shows that the younger is the
cluster, the brighter is the completeness limit required to
minimize the difference between SBF computed by including all
stars of the CMD, and SBF obtained by including only stars with $V
< V_{cut}$. For clusters as old as $t=4\, Gyr$, the uncertainty of
SBF introduced by incompleteness is $\leq 0.2\, mag$ in $\overline
V$, and $\leq 0.03\, mag$ in ($\overline V - \overline I$) if
$V_{cut}\geq 23\, mag$. For clusters younger than $\sim 900\, Myr$
the uncertainty of SBF keeps below $0.2\, mag$ in $\overline V$,
and $\lsim 0.05\, mag$ in ($\overline V - \overline I$) if
$V_{cut}\geq 21\, mag$.

As regards the LMC clusters considered here, the completeness of
NGC~2209 and $Hodge\, 14$ is shown to be larger than 90\% for $V$
and $I \leq 23$, while for younger clusters (NGC~1805, NGC~1818,
NGC~1831, NGC~1866, NGC~1868) it is larger than $\sim 80$--$90\%$
for $V$ and $I \lsim 21$ except for the innermost annulus, i.e.
$r\leq 3.6\arcsec$ (de Grijs et al 2002a, 2002c; Brocato et al.
2003). As for old clusters (NGC~1754, NGC~1916, NGC~2005, and
NGC~2019), we used the photometric data by Olsen et al. (1998),
who made deep exposures in both $F555W$ and $F814W$ filters. This
allows us to reach $V\sim 25\, mag$, well below the MS--TO point.
Again, for all these clusters completeness at $V \leq 23\, mag$ is
assured to be more than 80\% for $r\geq 5\arcsec$.

Thus, the uncertainty of $\overline V$ and $\overline I$ due to
the incompleteness is very small, well--below $0.2\, mag$ for all
the ages considered. This ensures that $\overline V$ and
$\overline I$ are derived with a degree of precision adequate for
the level of intrinsic uncertainty due to statistical fluctuations
expected for LMC clusters (Section~\ref{section:statistics}).

Crowding effects might also be relevant in evaluating the
numerator of Eq. \ref{eq:eqTS}. The technique of distributing
artificial stars also helps in studying this issue. From Fig. 2 by
de Grijs et al. (2002c) it is evident NGC~1831, NGC~1868, NGC~2209
and $Hodge\,14$ are not affected by crowding effects showing a
$\sim 100\% $ completeness for the 3 brightest magnitudes of the
cluster stars. A similar evidence can be derived for NGC~1866
(Brocato et al. 2003). On the other hand, NGC~1805 and NGC~1818
may suffer a $10\%$ of missed stars within $7.2\arcsec$ from the
center (representing less than $3.5\%$ of the PC area). Since the
completeness functions have a statistical meaning, a single young
cluster has been further analyzed in order to check whether very
bright stars are missed due to crowding effects and/or saturation
problems. For clusters younger than a few billion years, we must
pay a particular attention to bright and cool AGB stars, which may
strongly affect the SBF measurements. In order to check the
completeness of the brightest end of the LF, we retrieved for each
cluster the images and the $JHK_s$ photometry available in the
Final Release of the Two Micron All Sky Survey
(2MASS)\footnote{http://www.ipac.caltech.edu/2mass/}. Then, a star
by star cross--identification between the two sets of photometry
(HST and 2MASS) has been performed to avoid cool AGB stars within
the observed field could be missed in the final photometric list
used to derive SBF. If we missed one or more cool stars in the HST
photometry, their corresponding $V$ and $I$ magnitudes have been
obtained from the literature. The $V$ and $I$ magnitudes of such
bright and cool AGB stars are available in the literature, mainly
because of the large efforts done in the past in searching AGB
carbon--rich stars (C--stars) in the LMC clusters (e.g. Aaronson
\& Mould 1982, Westerlund et al. 1991).

However, there is no guarantee that such bright stars found within
the observed cluster field all belong to the cluster. For this
reason, we provide -- as an indicative uncertainty of the SBF
measurements -- half the difference between the SBF, computed by
using only stars within the PC field, and the SBF obtained by
adding to the photometric list the \emph{missed} bright stars.
This is a safe assumption, which probably leads to overestimate
the uncertainty. Nevertheless, the primary goal of this paper is
to explore the general behavior of optical SBF of young stellar
populations, while leaving a detailed and quantitative analysis on
LMC clusters to a forthcoming paper.

In the data analysis we also looked for another severe effect of
crowding, that is the blending effect. The completeness curves we
considered are corrected for it as well as for superposition of
multiple randomly placed artificial stars. However, to make a
further check, we estimated the number of expected blended pairs
as discussed by Stephens et al. (2001). Dealing with WFPC2/HST,
even in the worst case of a very densely populated core of a
cluster like NGC 1866, the number of blended pairs formed by two
identical giant stars is estimated to be of the order 0.0001\% for
the PC, and about 0.01\% for the WF chips.

The field stars contamination is not severe for the selected
clusters (with the exception of $Hodge\, 14$). The contribution of
field stars on SBF measurements is evaluated by comparing SBF
magnitudes derived from the whole region covered by the WFPC2 with
the results obtained from the PC area only, which typically
includes most of the stars of the cluster.

Finally, the four truly old clusters show SBF magnitudes in very
good agreement with the average SBF values obtained for the GGCs
by AT94. This last point can be also seen as \emph{a posteriori}
verification that our method of measuring SBF from high quality
and deep photometric data is reliable, at least for the purpose of
the present paper.

\subsection{Age Determination}
\label{section:Age_determination}

\begin{figure*}
\begin{center}
\plotone{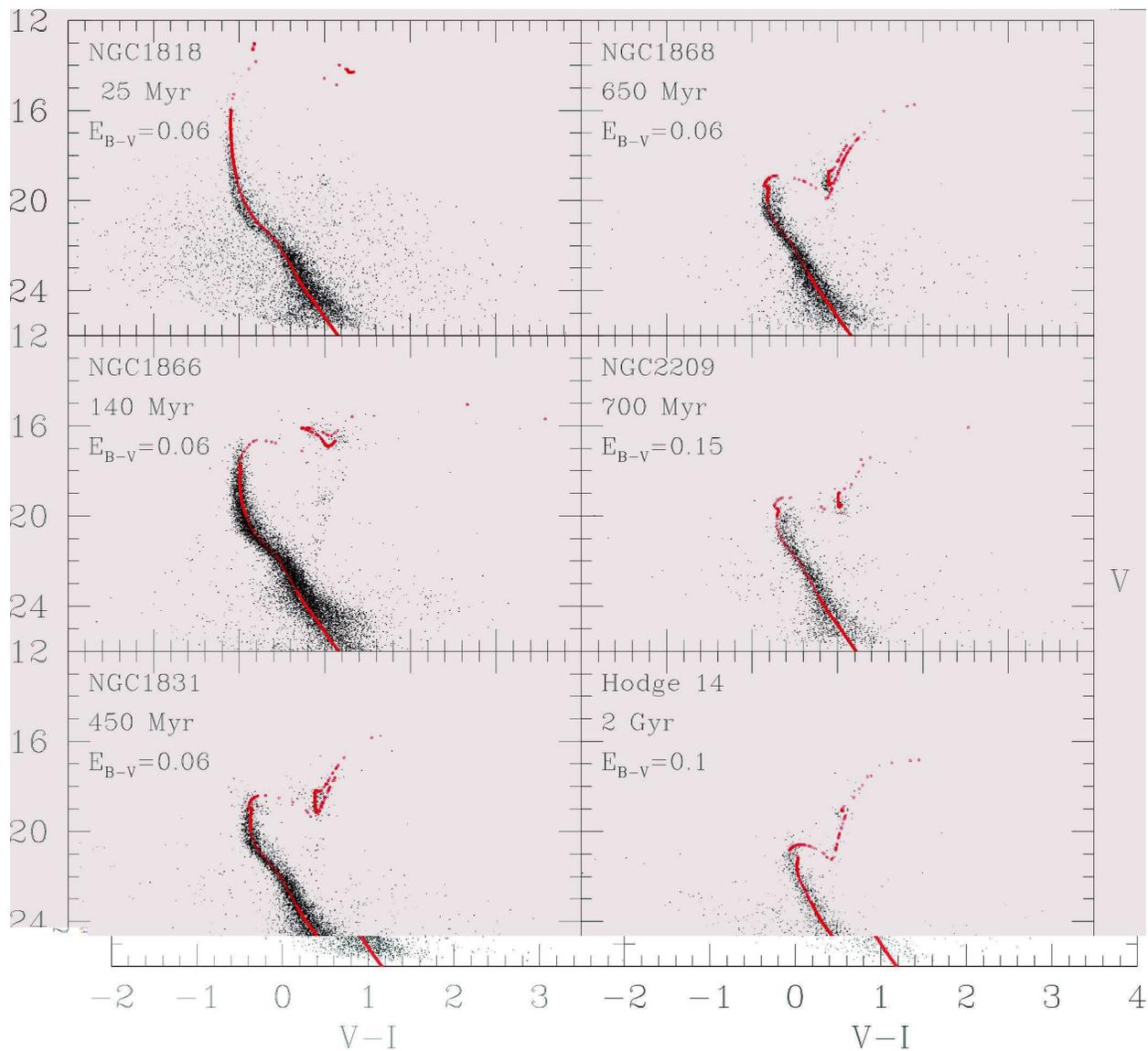} \caption{Observed (black dots) and synthetic (red
dots) CMDs of a sample of LMC clusters are compared. In each panel
the best fit is plotted, together with the derived reddening
value. $Z=0.008$ and an absolute distance modulus of $(m-M_V)_0 =
18.4$ are adopted. Photometric errors are also indicated on the
right side of each panel}. \label{fig:cmd_clusters}
\end{center}
\end{figure*}

\begin{figure*}
\epsscale{1} \plotone{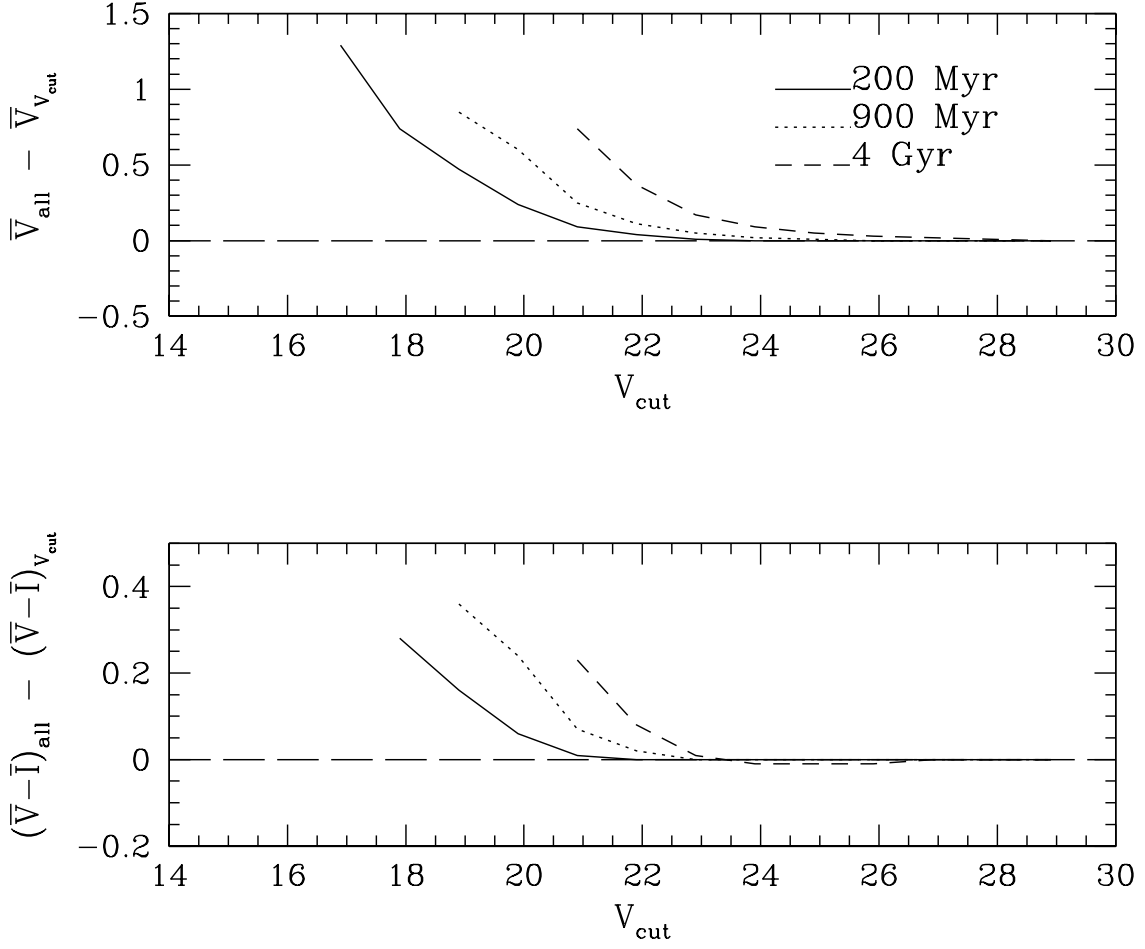} \caption{Impact of the completeness
limit $V_{cut}$ on SBF. The differences between SBF obtained using
a synthetic CMD complete down to star mass $m=0.1\, M_\sun$ and
SBF from the same CMD but excluding stars with $V>V_{cut}$ (i.e.
100\% of incompleteness at $V>V_{cut}$). The labelled ages and
fixed metallicity $Z=0.008$ are adopted.} \label{fig:sbfVcut}
\end{figure*}

We used present synthetic CMDs to consistently evaluate the age of
each cluster. In Fig.~\ref{fig:cmd_clusters} we show the
comparison between the observed CMDs and the synthetic ones. For
all the clusters, we assumed a metallicity equal to $Z=0.008$ and
an absolute distance modulus of $(m-M)_0 = 18.4\pm 0.1$. The
reddening value of each cluster is derived from the best--fitting
procedure as discussed in Walker et al. (2001). The assumed
distance to the LMC appears justified in the light of recent
distance measurements, which attempt to reconcile the long-- with
the short--distance scale \citep[e.g.][]{Salaris+03, Alcock+04}.

The cluster age and the related overall uncertainty is reported in
Table~\ref{table:table_lmc} (Col.\ 7). They are in good agreement
with ages listed by MG03. The interstellar reddening values fall
within the range measured for the LMC, e.g. $E_{B-V}=0.06$--0.20
\citep{Westerlund97}.

Fig.~\ref{fig:cmd_clusters} confirms the extremely high degree of
accuracy that our SSP models reach in simulating the CMDs of young
stellar populations. On the other side, we remark that our
procedure is fully consistent, since the same theoretical
framework was adopted for computing SBF, and for the CMD--analysis
aimed to obtain the age of each cluster.

\section{Models vs. Observations}
\label{section:sbf&LMC}
\subsection{Optical SBF }

\begin{figure*}
\epsscale{1}  \plotone{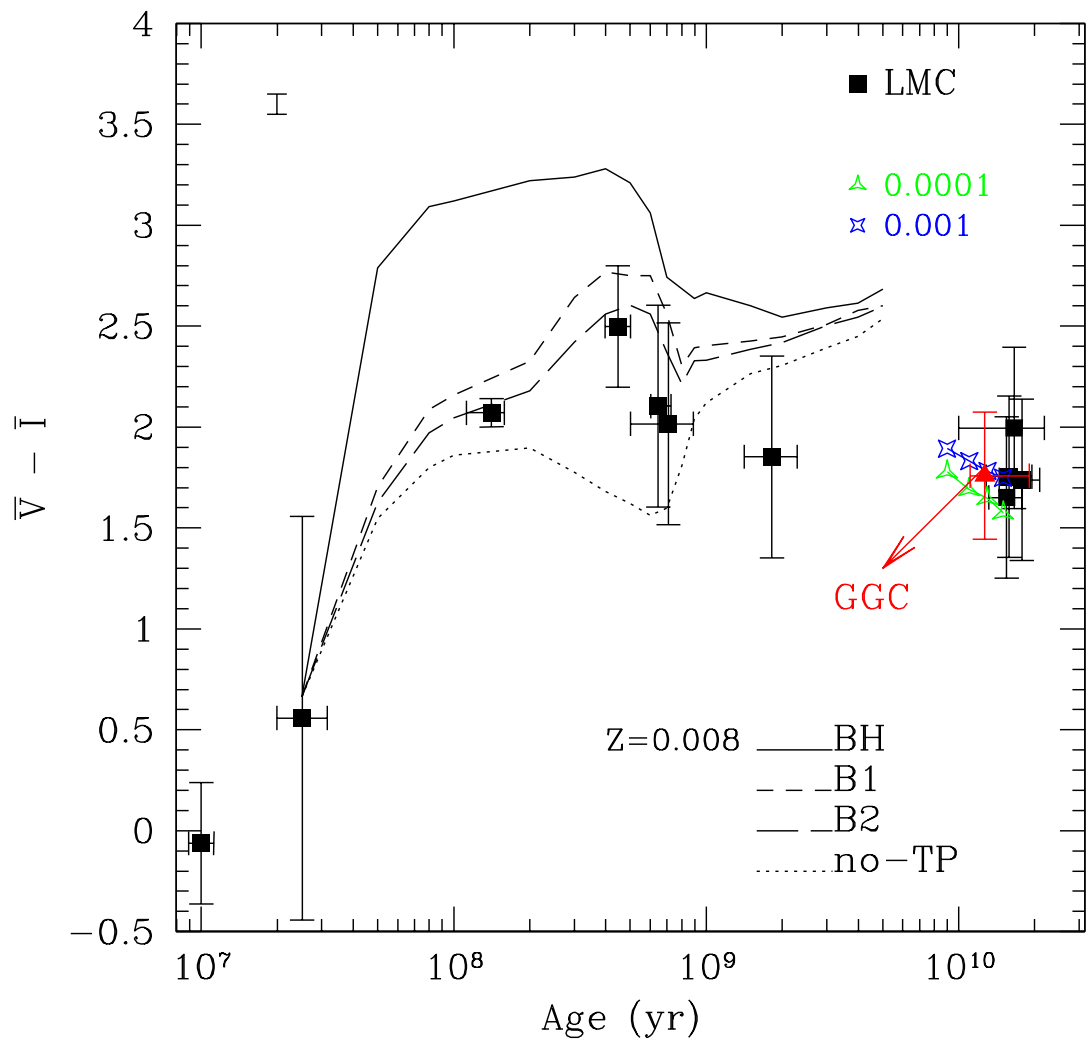} \caption{The $(\overline
{V}-\overline {I})$ fluctuation color vs. age. For ages lower than
$ 5\, Gyr$ models with $Z=0.008$, and different mass--loss rates
are plotted (symbols are as in Fig.~\ref{fig:sbfvsage}). For ages
larger than $5\, Gyr$ models with $Z=0.0001$ (green three--pointed
stars), and $Z=0.001$ (blue four--pointed stars) are from Paper I.
The LMC star clusters fluctuation colors are shown as black filled
squares. The red triangle refers to the mean $(\overline
{V}-\overline {I})$ color of GGCs (data from AT94). At the top
left side we report the mean error--bar of SBF models.}
\label{fig:sbf&LMC}
\end{figure*}

In this Section, we compare SBF predictions with the optical
measurements. As discussed in Section~\ref{section:sbfandtp}, the
($\overline V - \overline I$) fluctuation color is sensitive to
the mass--loss rate, active along the final stages of the AGB. In
Fig.~\ref{fig:sbf&LMC} observational data for clusters with age $t
< 5 \, Gyr$ are located within the two theoretical curves
representing models \emph{without} TP--AGB stars ($no$--$TP$,
dotted line), and models computed by assuming the BH mass--loss
rate (solid line). The cluster $Hodge\, 14$ is the only exception:
we shall discuss it later. Being the less efficient rate explored
in the present work, the BH mass--loss rate predicts a large
number of TP stars which are responsible for very 'red'
$(\overline {V}-\overline {I})$ colors of young SSPs. In spite of
the large error--bars of the LMC clusters data,
Fig.~\ref{fig:sbf&LMC} suggests that $B1$-- and $B2$--models give
a better fit than $BH$--models for nearly all the clusters. In
particular, NGC 1866, for which we succeed in minimizing the
uncertainties, is well fitted by models with mass--loss
prescriptions by B95.

In the same figure, old and very metal--poor models ($t>5 \, Gyr$,
$Z=0.001$ and 0.0001) published in Paper I are plotted as
three/four--pointed stars. They have been computed by assuming a
Reimers mass--loss rate (see Paper I for details). Note that,
models of age $t>5 \, Gyr$ fully consistent with the present
theoretical scenario will be discussed in a forthcoming paper (see
note 5).

\begin{figure*}
\plottwo{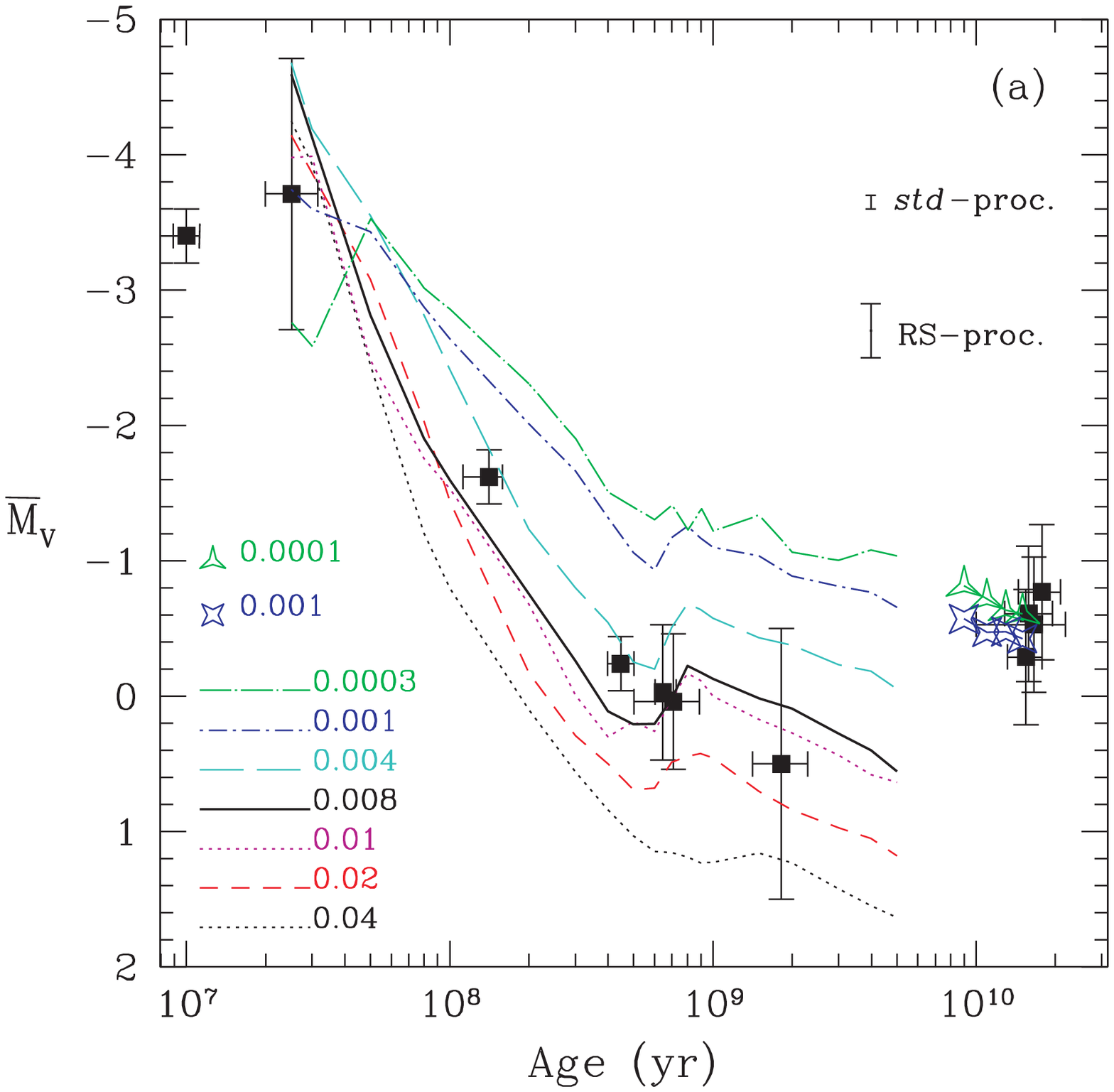}{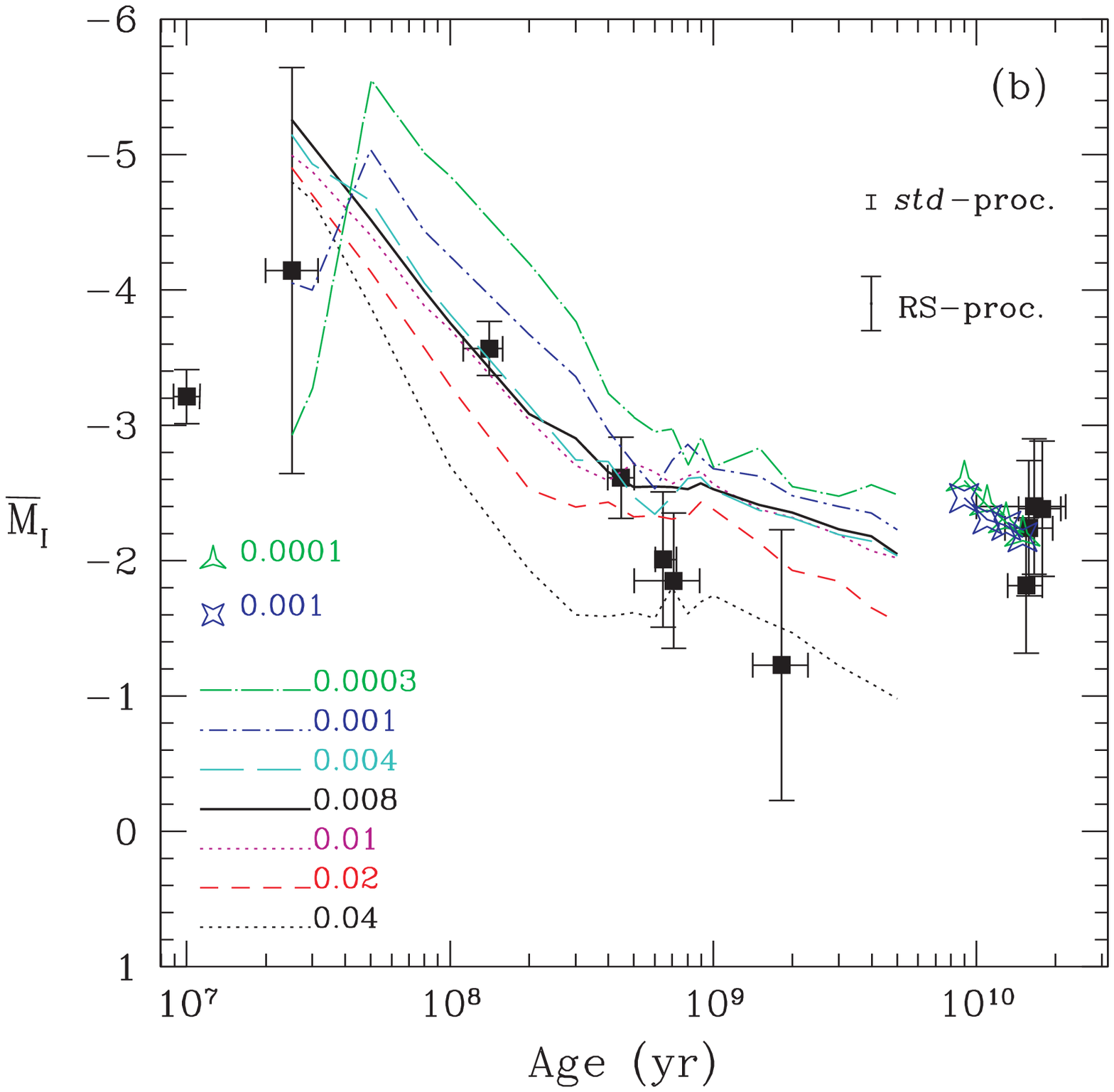} \caption{$\overline {M}_V$ (panel $a$)
and $\overline {M}_I$ (panel $b$) are plotted as a function of
age. For ages lower than $ 5\, Gyr$ models including $B1$
mass--loss rate are considered only. Models for $t>5\, Gyr$ are
from Paper I (symbols as in Fig.~\ref{fig:sbf&LMC}).
Black filled squares refer to the measured SBF
amplitudes for LMC star clusters. The expected theoretical
uncertainties are shown at the upper right corner of each panel
(see text).
} \label{fig:sbfVImags}
\end{figure*}

Fig.~\ref{fig:sbfVImags} exhibits the $\overline M_V$ and
$\overline M_I$ behavior with age. For $t< 5\, Gyr$ we plot SBF
models of all the metallicities presented in the present paper;
$B1$ mass--loss scenario is taken only. For older ages, models
with $Z=0.001$ and $Z=0.0001$ are taken from Paper I. The
theoretical SBF refer to the asymptotic values
(\emph{std}--procedure) for which the indetermination is of the
order of $0.05\, mag$ as already discussed. We also evaluated the
\emph{intrinsic} uncertainty due to stochastic effects on the
number of bright stars (\emph{RS}--procedure) from the models with
the faintest $M_V^{tot}$ (Table~\ref{table:table_sbf}). According
to the discussion in Section \ref{section:statistics}, the fainter
is the integrated $V$ magnitude of the cluster, the larger is the
\emph{intrinsic} uncertainty. We finally find out that it is of
the order of $0.2 \, mag$.

The general trend of $\overline M_V$ for LMC clusters is well
reproduced by models in the explored age range
(Fig.~\ref{fig:sbfVImags}$a$). The SBF measurements of clusters
younger than $5\, Gyr$ appear in agreement with models of
metallicities $Z=0.004-0.01$, which are appropriate for young and
intermediate LMC clusters \citep[e.g.][]{Westerlund97,
Mackey&Gilmore03}. SBF measurements for very old clusters are
fitted by 12--15 $\, Gyr$ with a lower metallicity models.

For $I$-band SBF (Fig.~\ref{fig:sbfVImags}$b$) the agreement is
still good in the case of measurements with small uncertainties.
Some relevant discrepancies arise for NGC 1868, NGC 2209, and
$Hodge\, 14$. Let us recall that the three quoted clusters have
the faintest integrated light in our sample, thus statistical
effects may be not negligible as inferred from
Fig.~\ref{fig:statistics1} (Section~\ref{section:statistics}). We
also remind the reader that, in order to avoid contamination by
field stars, in the case of $Hodge\, 14$ a small fraction (i.e. a
small mass) of the cluster is being analyzed only. This indication
is confirmed if detailed models are computed by assuming exactly
the integrated magnitude, and age reported in
Table~\ref{table:table_lmc} (Col.\ 3, 7). The $I$--band SBF,
derived by applying the $RS$--procedure ($\overline {M}_I^{RS}$)
provide values which are lower than $\overline {M}_I^{std,asym}$
and closer to the observed ones. In fact, for NGC 1868, NGC 2209,
and $Hodge\, 14$ we find respectively $\overline {I}^{RS}=16.6 \pm
0.5 $, $16.4 \pm 1.0$, and $16.5 \pm 0.9 $, which can be compared
to the corresponding asymptotic values $\overline {I}^{std,asym} =
15.93\pm 0.05$, $15.84\pm 0.05$ and $16.06\pm0.05$. Note that
similar computations performed for the massive cluster NGC 1866
give $\overline {I}^{RS}=15.0 \pm 0.1$. SBF in the $V$--band show
a similar behavior, in the sense that the agreement gets even
better if $\overline {M}_V^{RS}$ values are compared to
observations. Statistical effects are mainly driven by
fluctuations in the number of giant stars, then for
intermediate/old age populations they affect the $I$ and NIR bands
more than the optical ones (Fig.~\ref{fig:statistics1}).

Even with the present large error--bars and the limited sample of
LMC clusters we are dealing with, we can reach the following
conclusions:

1. SSP models including TP stars reproduce optical SBF of LMC
clusters reasonably well. This rules out the extremely high
mass--loss rate, thus driving stars to an early departure from the AGB;

2. A mild mass--loss rate (BH) appears inadequate, because too much
TP stars are foreseen. SSPs predict a very 'red' $(\overline
{V}-\overline {I})$ color, which is not supported by the observed
SBF values.

3. $B1$-- and $B2$--models seem to predict a number of TP stars
which can reproduce the SBF of the selected sample of LMC
clusters.

Due to the relatively small number of LMC clusters included in our
sample, the previous conclusions must be regarded as important,
though not conclusive, indications. Further observational
efforts are requested, both in improving the size of the sample and
in minimizing the uncertainties of the measurements.

\subsection{NIR SBF }
\label{section:obs_NIR}

G04 recently provided \emph{JHK} SBF--measurements of eight
'super--clusters' as obtained from 2MASS observations of several
LMC and Small Magellanic Cloud (SMC) stellar clusters. Each
super--cluster groups clusters within a given range of the
s--parameter (Elson \& Fall 1985), and corresponds to a SWB class
(Searle et al. 1980). The SBF derived from these super-clusters
have the remarkable advantage of relying upon a large number of
bright stars.  For this reason, we compared present models with
those NIR data.

Differently from Gonzalez and collaborators, who used the Cohen
(1982) ages for the SWB classes, we adopted the age calibration of
the s--parameter from two more recent works: Elson \& Fall (1988)
and Girardi et al. (1995). They provide similar results (within
$5-10 \%$) even if one is based upon \emph{canonical} stellar models
($log\ t = 6.05 + 0.079 s$: Elson \& Fall 1988) and the other one
on \emph{overshooting} stellar models ($log\ t = 6.227 + 0.0733
s$: Girardi et al. 1995).

\begin{figure*}
\center
 \includegraphics[width=7cm]{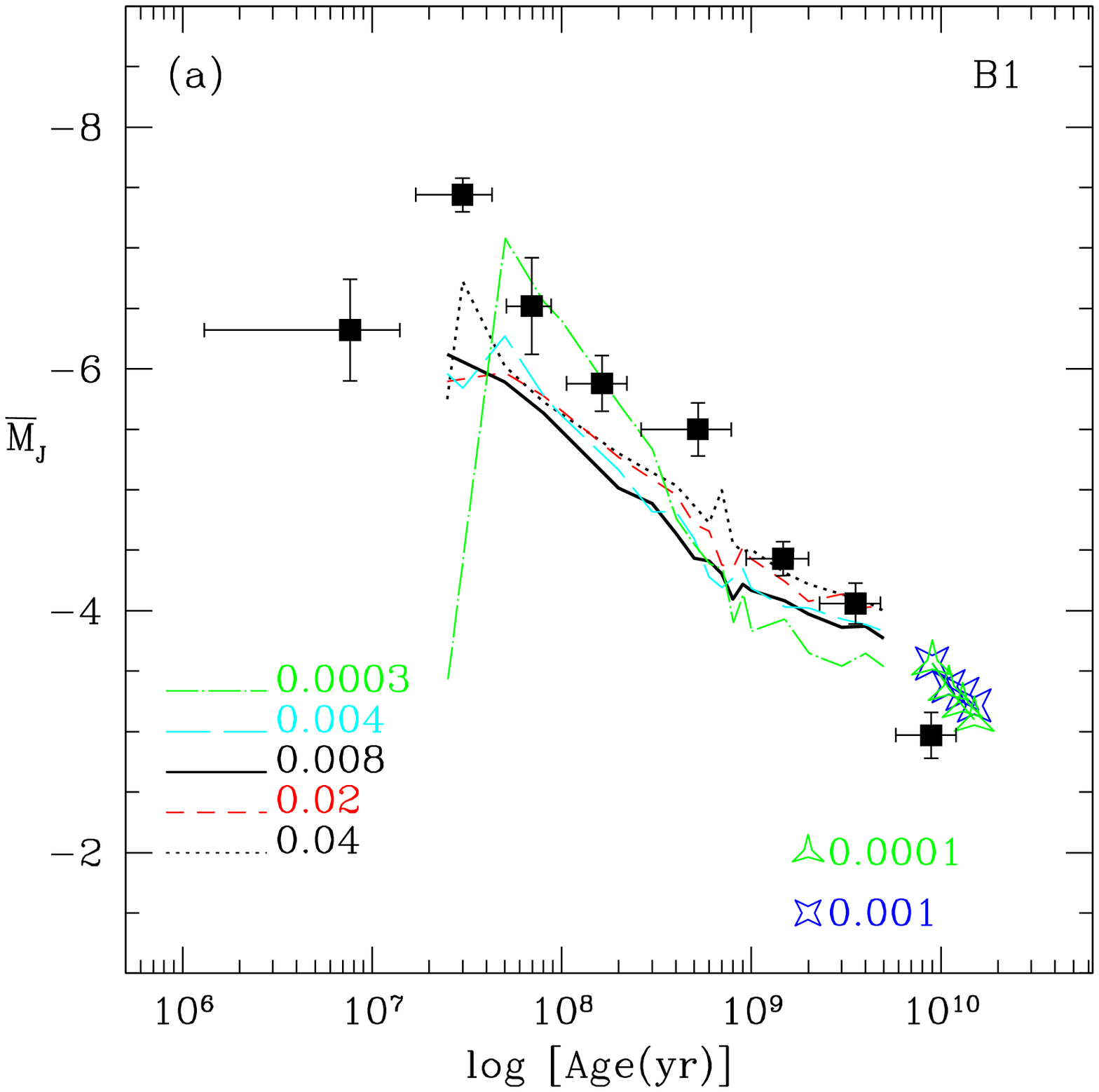}
 \includegraphics[width=7cm]{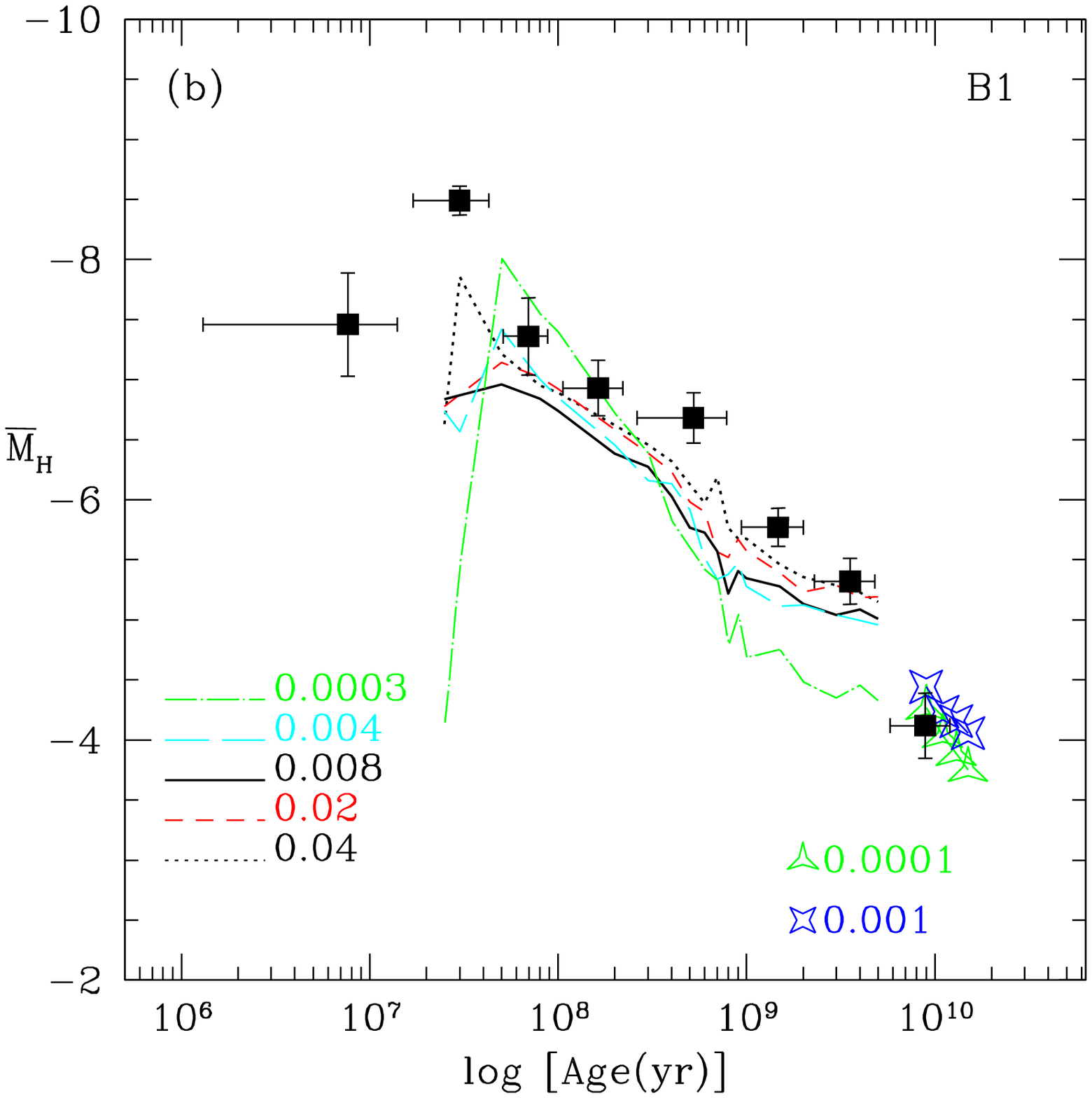}
 \includegraphics[width=7cm]{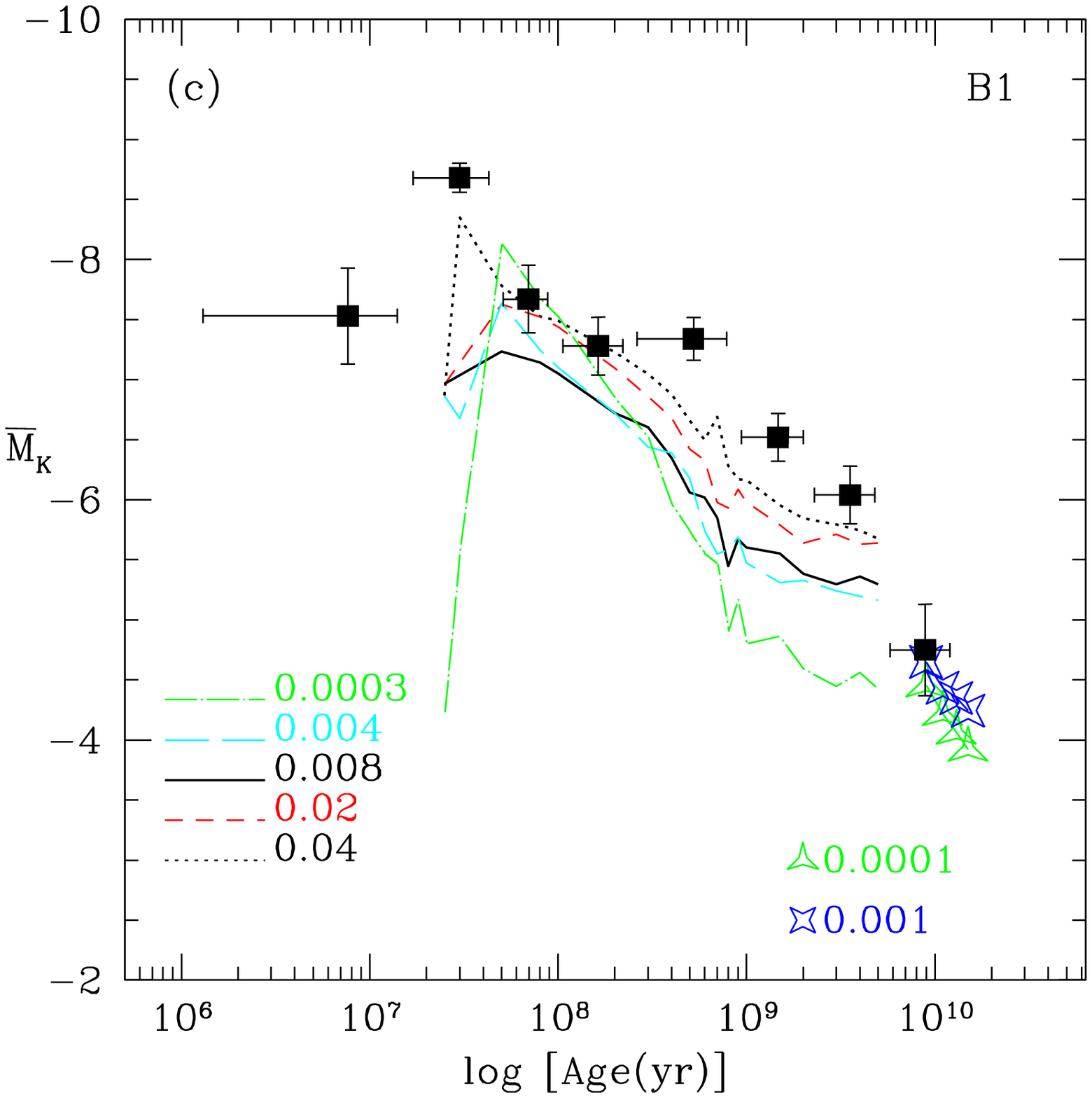}
\caption{$JHK$ SBF predictions of $B1$--models for selected values
of metallicity as a function of age. Models for $t>5\, Gyr$ are
from Paper I (symbols as in Fig.~\ref{fig:sbf&LMC}). The SBF of MC
super--clusters by G04 are plotted as filled squares. Age
uncertainties cover the interval spanned by each SWB class
according to the s--parameter values included in the
super--clusters. } \label{fig:sbfNIR}
\end{figure*}

In Fig.~\ref{fig:sbfNIR} we plot our \emph{JHK} SBF predictions as
a function of age for different metallicities, and the
$B1$--models. The observational data published by G04 (as
corrected after Gonzalez et al. 2005) are reported as filled
squares, with the related ages according to the \emph{canonical}
calibration. The age error--bars refer to the ages corresponding
to the initial and final s--parameter values of the clusters
included in each super-cluster. We re--scaled G04 measurements to
the LMC distance adopted in previous Sections. The models are
calculated for slightly different \emph{K}--band filter than 2MASS
\emph{K}--band filter ($K_s$), but the differences are negligible
for our present purpose \citep{Carpenter01}.

The general behavior of \emph{JHK} SBF data is reproduced by
models in all NIR bands. Both models and data show a jump around
$30-50\, Myr$ corresponding to the appearance of red and bright
AGB stars. After that, the luminosity of the AGB--tip decreases
with age, and the SBF NIR data decline accordingly. In spite of
the qualitative agreement concerning the general trend, the
quality of the fit is \emph{not} satisfactory.

The comparison shows that SBF from $B1$--models are systematically
fainter than NIR SBF magnitudes of the MC super--clusters. In the
range from a few hundreds Myr to few Gyr, only super--solar SSP
models give SBF magnitudes as bright as G04 data. It is well known
that MC clusters have chemical abundances lower or at most equal
to the solar value, so that metallicity variations do not appear
to properly solve the quoted discrepancy. Other models (G04) also
require unlikely high values of metallicity to reproduce the SBF
measurements obtained for super--clusters.

\begin{figure*}
\plottwo{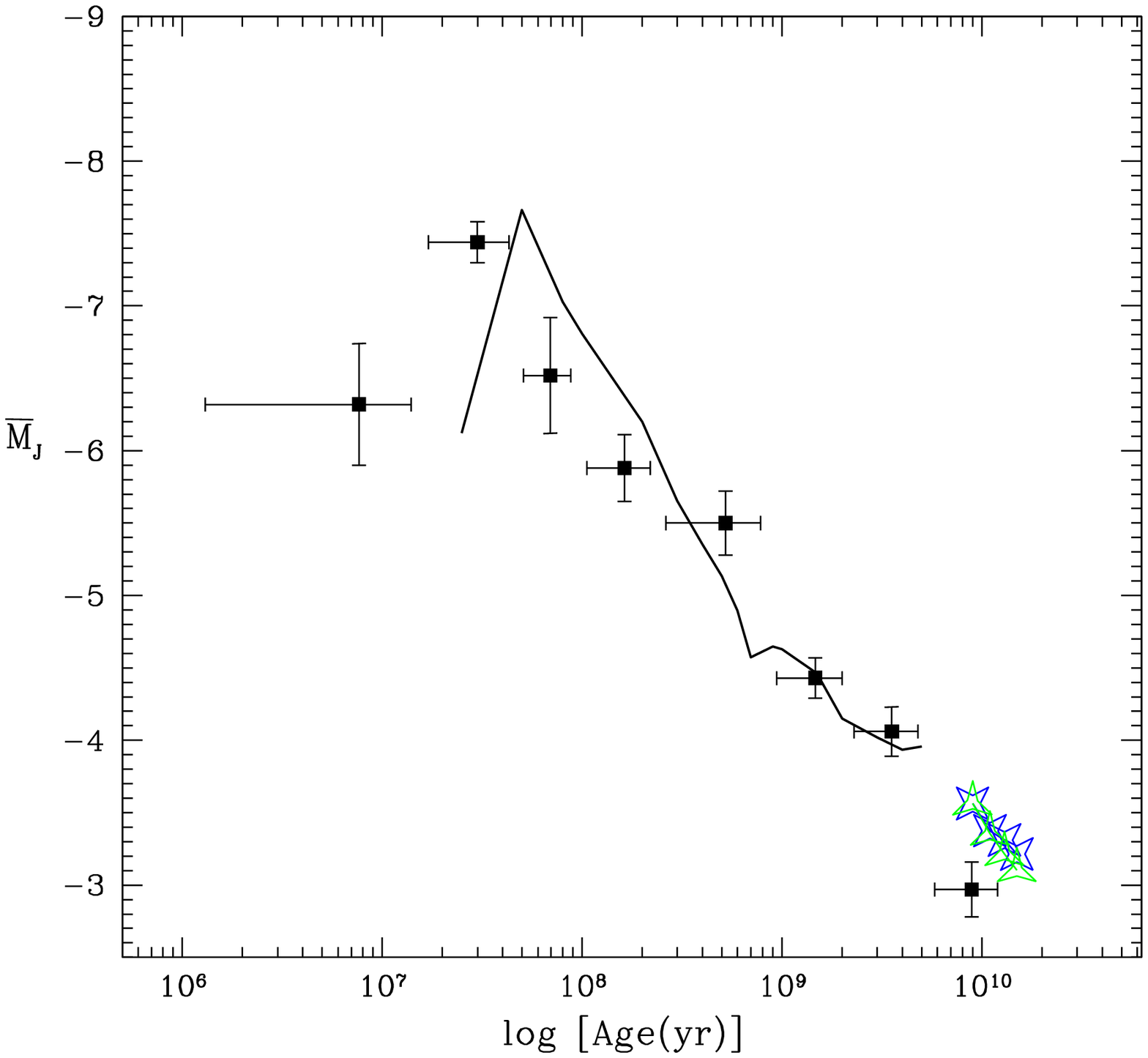}{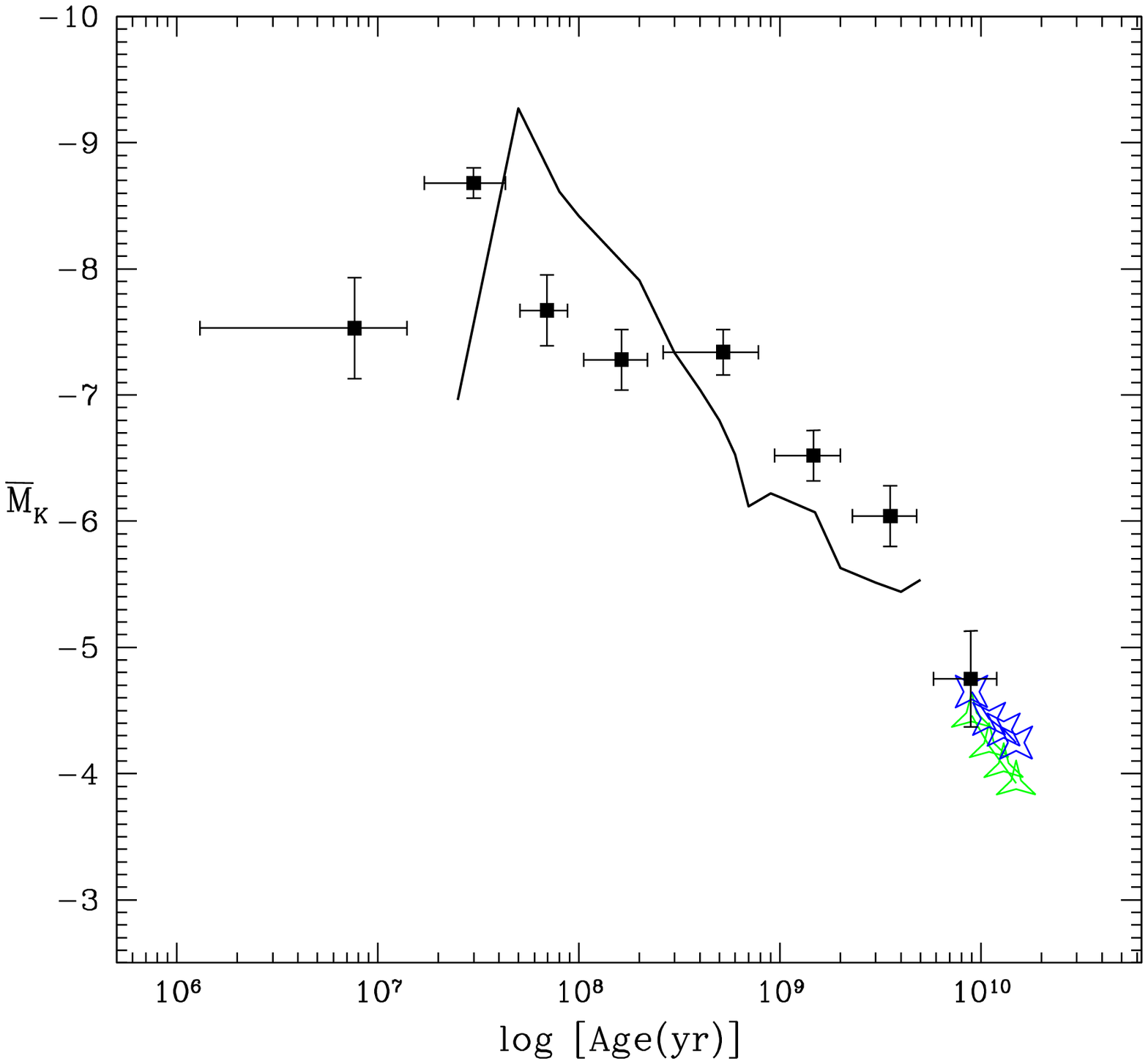} \caption{$J$-- and $K$-- SBF
predictions obtained by $BH$--models and metallicity $Z=0.008$
(solid black line). Models for $t>5\, Gyr$ are from Paper I
($Z=0.0001$ green three--pointed stars; and $Z=0.001$ blue
four--pointed stars). See the electronic edition of the Journal
for a color version of the figure. } \label{fig:sbfNIRBH}
\end{figure*}

We evaluated the impact of increasing the number of cool bright
AGB stars by comparing the super--cluster data to the SBF
predictions obtained by $BH$--models. Fig.~\ref{fig:sbfNIRBH}
(left panel) shows that \emph{J}--band SBF magnitudes seem to be
well fitted by $BH$--models with reasonable metallicity values.
Unfortunately, these models fail in reproducing \emph{K}--band
measurements, since the theoretical SBF amplitudes are brighter
than the data for $t \sim 100 Myr$ and fainter for older ages
(Fig.~\ref{fig:sbfNIRBH}, right panel). Hence, this scenario seems
to be ruled out, too.

Which kind of stars are missed in the models and what place
should they occupy in the CMD in order to move the theoretical NIR SBF
magnitudes to the observed values? Note that SBF predictions in
the optical bandpasses should not be affected by including that
kind of stars, because the SBF in that range of wavelength can
fit observations. This leads to investigate the lack of cool
bright stars in our models. In order to reproduce exactly the SBF of
super--clusters, we performed numerical experiments by assembling
the number of simulations needed to form a super--cluster with a
total mass ${\cal M} \sim 4\cdot 10^{6} \, M_{\sun}$, representing
an average value from classes II to VII of G04 at a given age
($B1$--models and $Z=0.008$). Obviously, the SBF derived for these
theoretical super--clusters resemble -- within the uncertainties
-- the SBF reported in Table~\ref{table:table_sbf}, because we are
dealing with asymptotic values.

\begin{figure*}
\plottwo{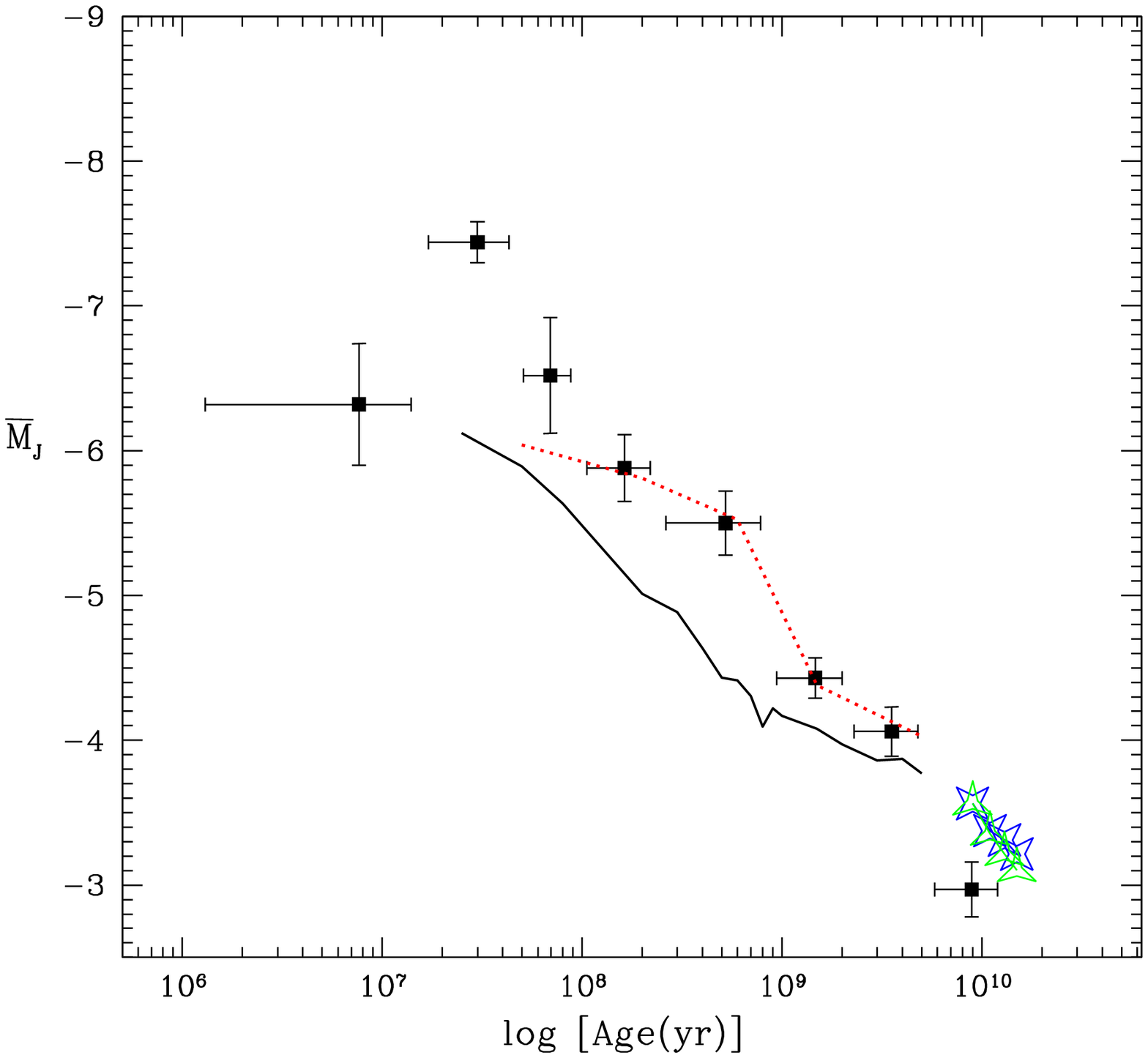}{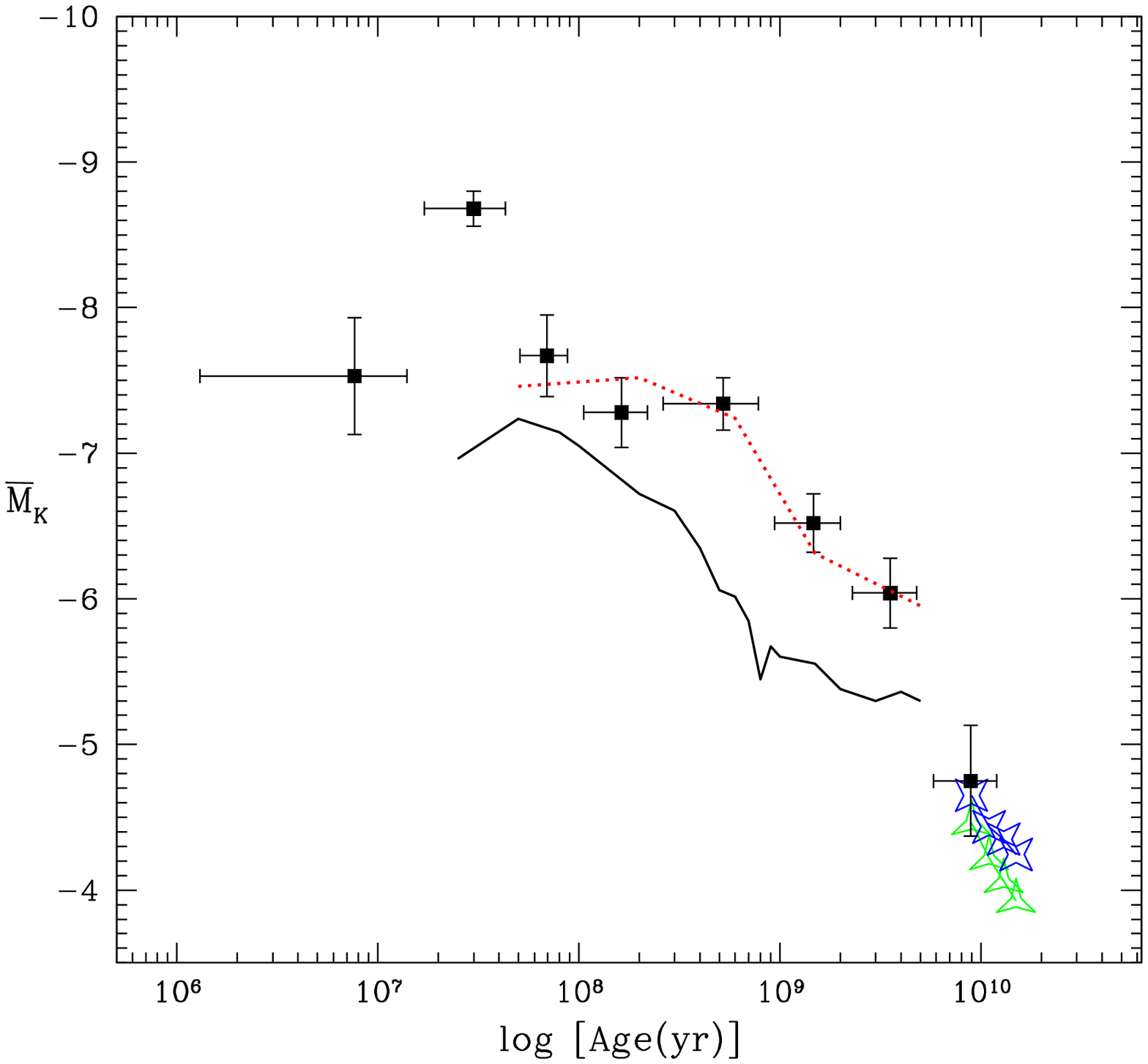} \caption{$J$-- and $K$-- SBF
predictions for theoretical superclusters obtained from
$B1$--models by including a contamination of M--type field stars
\emph{and} by artificially increasing the number of stars with
$J-K \gsim 1.3-1.4$ (C--stars ?) in the SSPs (red dotted line). As
reference original $B1$--models with $Z=0.008$ are reported (black
solid line). Symbols are as in Fig.~\ref{fig:sbfNIRBH}. See the
electronic edition of the Journal for a color version of the
figure.} \label{fig:sbfNIRfield}
\end{figure*}

As a first approximation, we verify what happens to theoretical NIR
SBF if a contamination of LMC supergiant M--type field stars is
included \citep{Nikolaev&Weinberg00}. Fig.~3 by G04 shows that a
non negligible number of bright stars at $K_s \simeq 10$ and
$J-K_s\simeq 1$--1.2 are present in the super--clusters
corresponding to I--IV SWB classes. If such a small contamination
of field stars is included into our theoretical super--clusters, we
find that models shift towards a higher luminosity. Furthermore, the
presence of NIR--bright AGB stars displaying $^{12}C$ enrichment,
due to the third dredge--up (C--stars), also affects SBF in these
bands. The $K_s$ $vs.$ $J-K_s$ CMDs of the super--clusters (Fig.~3
of G04, SWB classes IV, V and VI) show stars with $J-K \gsim
1.3-1.4$, the typical value used to select C--stars
photometrically in the SMC and LMC (Cioni et al. 2001; Cioni et
al. 2003, Raimondo et al. 2005). Cohen (1982) already stated that
these classes are "precisely those where integrated light (Persson
et al. 1983) and searches among the brightest red stars (Frogel \&
Cohen 1982; Aaronson \& Mould 1982) have revealed the presence of
luminous carbon stars.". If this is the case, the discrepancy
between NIR data and models should be reduced by moving from
$K$-- to $J$--band, and is expected to become negligible in the
optical range. This is exactly what happens -- as shown in
Fig.~\ref{fig:sbfNIR}($a,b,c$). From Fig.~3 of G04, we note that
SWB classes V and VI show a narrow giant branch with a group of
stars redder than $J-K \gsim 1.4$, roughly 10 and 4 for SWB
classes V and VI, respectively. By adding these stars to our
theoretical super--clusters, SBF predictions go up to the same
position of observational values (Fig.~\ref{fig:sbfNIRfield}).

All in all, these numerical experiments lead to the conclusion
that models with the proper MC metallicity can be reconciled with
observed NIR SBF for MC super--clusters if: 1) a field
contamination of M--type stars is assumed, \emph{and} 2) the
number of stars with $J-K \gsim 1.3-1.4$ is increased in the SSP
models.

In spite of the fairly good agreement obtained in
Fig.~\ref{fig:sbfNIRfield}, we explored a further possibility. Let
us recall that super--clusters are not really SSPs, being the
results of a sum of individual stellar clusters with slightly
different ages and chemical compositions. In addition, the
membership of an individual cluster to a given SWB class may be
uncertain (Girardi et al. 1995). Thus, we investigated the
possibility that super--clusters contain a certain fraction of
populations younger than the minimum age assigned to the
corresponding SWB classes. This suggestion is supported by the
fact that CMDs of I, II, III, and IV SWB classes in Fig.~3 by G04
exhibit features which could be related to the presence of
populations with chemical compositions and/or ages different from
the corresponding SWB class (see for example the bi--modal red
giant branches).

\begin{figure*}
\plottwo{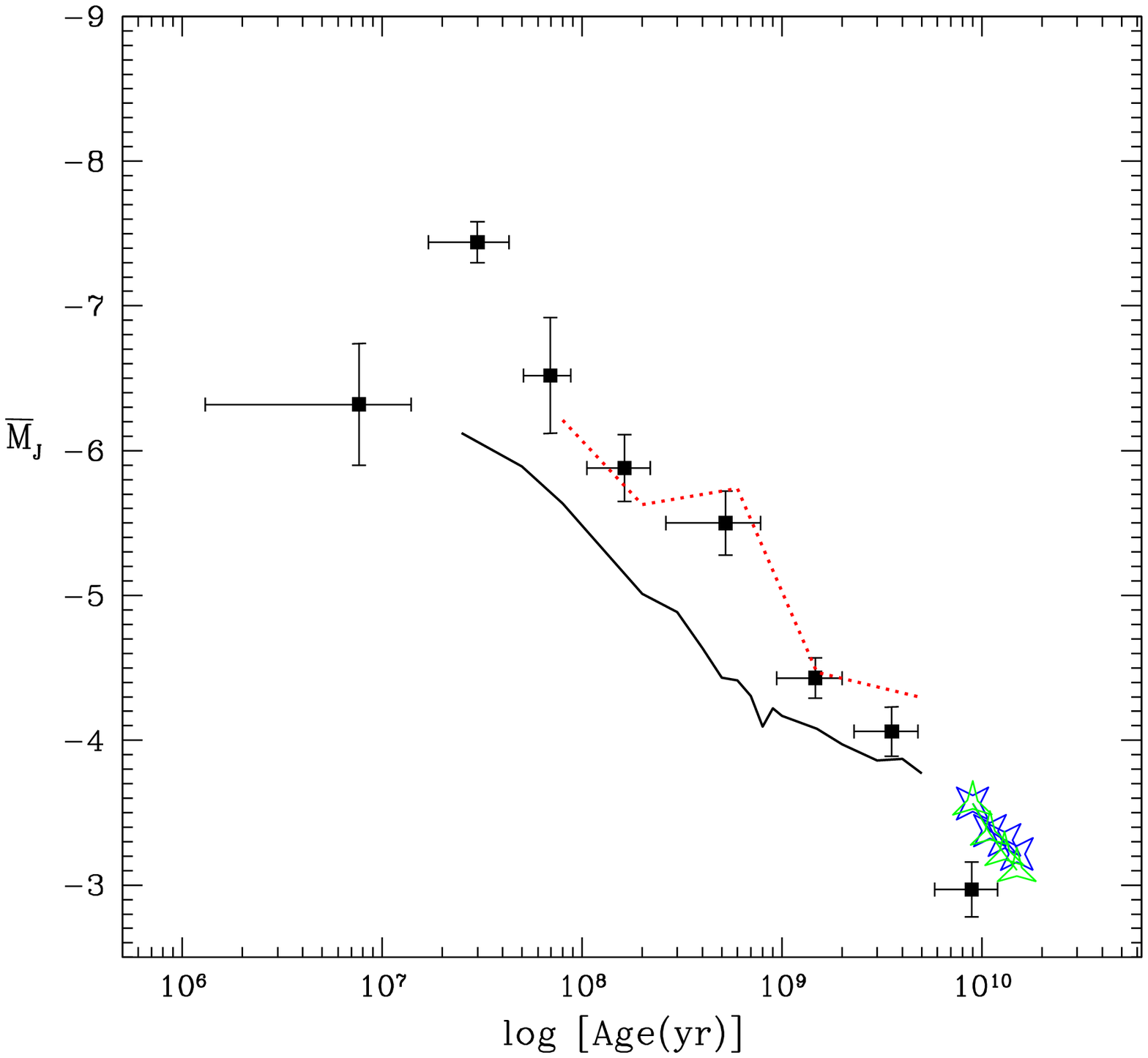}{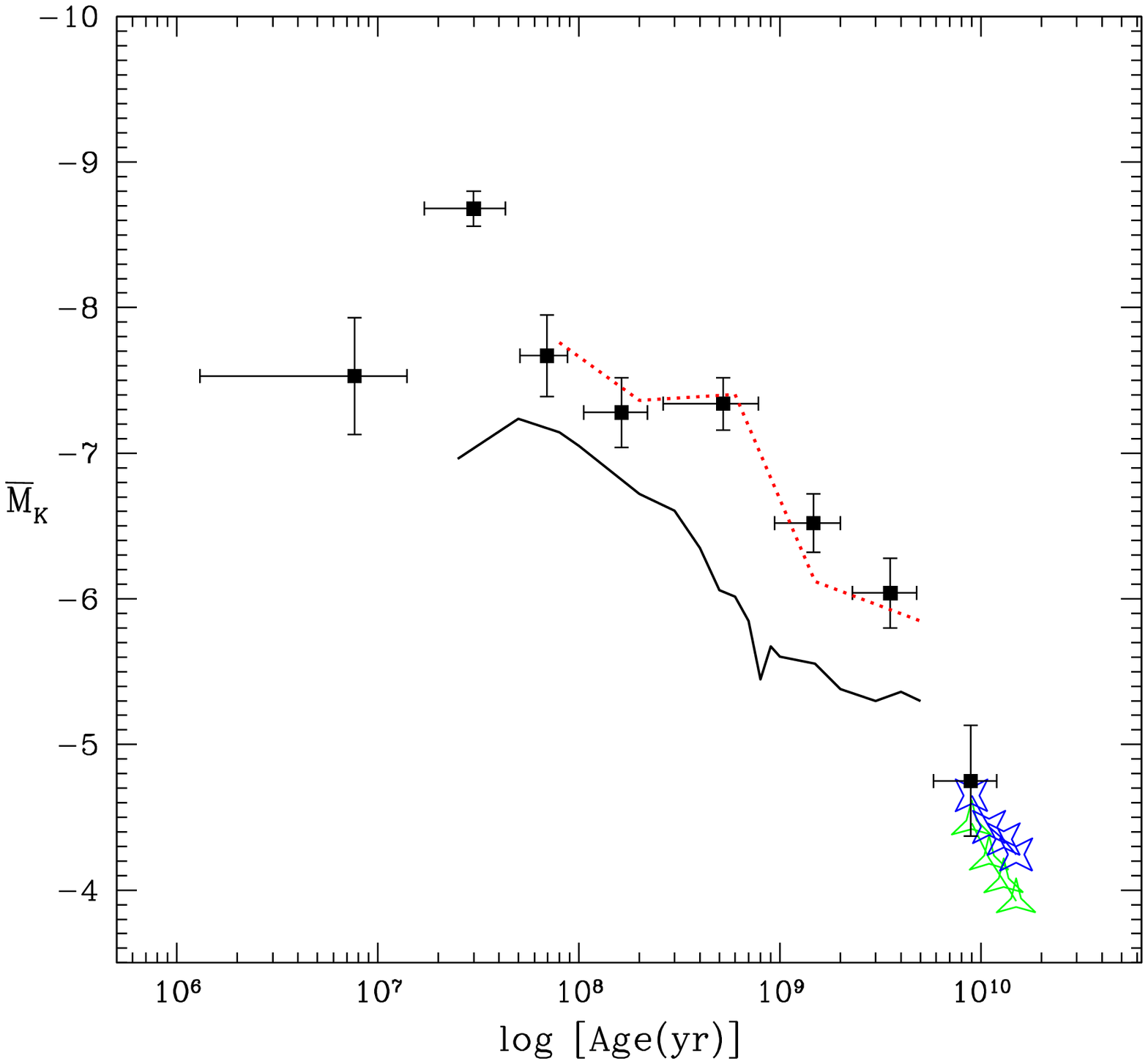} \caption{$J$-- and $K$-- SBF for
composite stellar populations (red dotted line, see text). Symbols
are as in Fig.~\ref{fig:sbfNIRBH}. See the electronic edition of
the Journal for a color version of the figure.}
\label{fig:sbfNIRcomp}
\end{figure*}

Numerical experiments performed for SWB classes from II to VI show
a relevant increment of the SBF magnitudes if a stellar population
younger than the typical age value assigned to the corresponding
SWB class is included. In particular, we find that the presence of
one (or few) younger cluster leads to predict SBF in agreement
with observations (Fig.~\ref{fig:sbfNIRcomp}). Keeping in mind
that the results in Fig.~\ref{fig:sbfNIRcomp} do not represent the
only solution, we note that the percentage of the young population
required to predict the SBF NIR data is of the order of 10\%. In
other words, if one (or few) cluster is included in a given SWB
class because of its s--parameter is overestimated (for whatever
reason, see Girardi et al. 1995), this leads to brighten the SBF
amplitudes for that SWB class. On the contrary, numerical
experiments show that an underestimation of the s--parameter of
one (or few) cluster would not significantly affect the
super--cluster SBF measurements.

In conclusion, we showed that NIR SBF of the super--clusters by
G04 can be reproduced -- though not in a definitive way.
Further observational and theoretical efforts are required to
improve the understanding of NIR SBF amplitudes of young and
intermediate age stellar populations.

\section{Summary and Conclusions}
\label{section:conclusions}

We presented new theoretical SBF amplitudes for single--burst
stellar populations of young and intermediate age ($25\, Myr \leq
t \leq 5\, Gyr$) and  metallicity ranging from $Z=0.0003$ up to
0.04. The new SBF models are based on an updated version of the
stellar population synthesis code already used to derive SBF for
old stellar populations in Paper I. In the present version of the
code, we used the recently published evolutionary--tracks database
of Pietrinferni et al. (2004), and $T_{eff}$--colors relations
from Westera et al. (2002). A particular care has been paid to the
simulation of the properties of intermediate--mass AGB stars
experiencing the TP--phase. The time evolution of core mass, and
luminosity, together with the overall evolutionary time--scale of
these stars have been evaluated according to prescriptions by
WG98. Additionally, the number of TP--AGB stars is also triggered
by mass--loss efficiency. Therefore, in order to evaluate the
impact of this type of stars upon fluctuation amplitudes, three
different mass--loss scenarios were simulated ($BH$, $B1$, and
$B2$), along with the extreme case of no TP--stars at all.

The resulting new database of stellar population models covers a
wide range of chemical compositions and ages. The accurate SSP
models allowed us to successfully fit the observed CMD features of
a sample of LMC star clusters imaged with the WFPC2/HST ($V$ and
$I$ bands). Age, metallicity, and reddening for all clusters are
derived and successfully compared with literature estimations.

Owing to the Monte Carlo technique, which is the basis of our
method to derive fluctuation amplitudes, we estimated the cluster
integrated magnitude $M_V^{Tot}$ required to minimize
indetermination caused by stochastic effects due to random
variation of the number of bright stars affecting SBF
measurements. We find that the procedure used to compute SBF from
individual stars of a synthetic CMD (\emph{RS}--procedure)
provides --within uncertainties-- the same results as the
\emph{std}--procedure if $M_V^{Tot} \lsim M_V^{Tot,max}$, being
$M_V^{Tot,max}$ a function of the stellar population age and
metallicity. This has a direct application in the observational
field. Firstly, once the absolute integrated magnitude of the
measured sample of stars is known, the \emph{RS}--procedure
provides a tool to evaluate the intrinsic uncertainty of SBF
measurements as derived by individual stars photometry of a real
stellar system (Fig.~\ref{fig:statistics1}). Secondly, the SBF
derived from the photometry of spatially resolved systems can be
compared properly with the theoretical SBF asymptotic values only
if the stellar system integrated magnitude satisfies the condition
$M_V^{Tot} \lsim M_V^{Tot,max}$, otherwise stochastic effects
prevent a reliable and meaningful comparison.

By focusing the attention on optical SBF, we performed the first
$V$ and $I$ bands SBF measurements for 11 LMC cluster by using the
WFPC2/HST photometry of individual stars. The explored age ranges
from $\sim$10$\, Myr$, for the very young cluster NGC\ 1805, up to
the typical age of Galactic globulars (NGC\ 1754, NGC\ 1916, NGC\
2005, NGC\ 2019). The comparison of SBF measurements with our
models showed a good agreement with both observed fluctuation
magnitudes and colors if metallicities of $Z= 0.008$--0.01 and
$Z=0.001$--0.0001 are adopted respectively for young/intermediate
age and very old clusters. The $(\overline {V}-\overline {I})$
fluctuation color has been found to be sensitive to the adopted
mass--loss scenario along the TP--phase. The comparison between
SBF models and measurements suggests that B95 mass--loss rates
better simulate the observed LMC clusters fluctuation colors.

Few interesting features have been identified in the
time--evolution behavior of fluctuation magnitudes and colors of
the present models. Sizable variations arise at $t \sim 50\, Myr$
and $t \sim 1\, Gyr$. These jumps correlate with the RGB-- and
AGB--phase transitions. For stellar populations younger than $1\,
Gyr$ the high sensitivity exhibited by SBF magnitudes to age
variation supports the use of the SBF tool to evaluate ages of
young stellar clusters in Local Group galaxies.

The capability of the stellar population synthesis code of
directly managing parameters and physical processes characterizing
the TP--AGB evolutionary phase allows us to analyze in depth the
SBF dependence on TP--AGB stars. It is worth remarking that the
different mass--loss scenarios affect both SBF magnitudes and
colors for populations older than $1\, Gyr$ only for few tenths of
magnitudes, confirming results given in Paper I. On the other
side, for $t< 1\, Gyr$ the SBF amplitudes appear to be highly
dependent on the adopted mass-loss scenario. Therefore, in this
age range SBF can be used to infer information on the mass-loss
efficiency of both resolved and unresolved young stellar
populations.

The comparison with NIR SBF observations of MC super--clusters has
shown that our models reproduce the general trend of the data but
a deeper analysis discloses that unlikely high values of
metallicity are required to fit the super--clusters data. We have
shown that composite stellar populations or contamination by field
stars coupled with a more precise simulations of very cool stars
(C--stars) may reconcile NIR SBF of MC super--clusters with
theoretical predictions.

Therefore, SBF studies in both the optical and NIR regime evidence
that further theoretical and observational efforts are needed to
improve the models accuracy, and the reliability of measurements
for nearby well known objects. On the other side, SBF should be
regarded as a very valuable tool to improve our knowledge of
unresolved stellar populations in distant galaxies for which SBF
measurements can be used to detect the presence of young and
intermediate age stellar populations, and to investigate their
evolutionary properties more efficiently than "classical"
integrated light studies.

\begin{acknowledgments}

We thank Dr. R. Gonz\'alez-L\'opezlira for providing constructive
and insightful comments that have greatly improved the paper. It
is a pleasure to acknowledge Adriano Pietrinferni and Santi
Cassisi for providing evolutionary tracks, and S. Shore for useful
discussions and for a careful reading of the manuscript. Financial
support for this work was provided by MIUR--Cofin 2003. This work
made use of computational resources granted by the Consorzio di
Ricerca del Gran Sasso according to the Progetto 6 {\it 'Calcolo
Evoluto e sue Applicazioni (RSV6)'}--Cluster C11/B. This paper is
based on observations made with the NASA/ESA Hubble Space
Telescope, obtained from the ESO/ST-ECF Science Archive Facility,
and data products from the Two Micron All Sky Survey, which is a
joint project of the University of Massachusetts and the Infrared
Processing and Analysis center/California Institute of Technology,
funded by the National Aeronautics and Space Administration and
the National Science Foundation.

\end{acknowledgments}

\appendix
\section{Modelling the TP--AGB stellar evolution phase}
\label{section:appendix1}

The present SSP models rely on the evolutionary--tracks database
by P04, which include all the most updated physical inputs in
computing stellar models, as for example the new equation of state
by Irwin et al. (2004), and recent nuclear reaction rates by
Angulo et al. (1999), and Kunz et al. (2002) for the
$^{12}C(\alpha, \gamma)^{16}O$ (see P04 for more details). The
database covers a wide range of chemical compositions and stellar
masses. Intermediate and low mass stellar models end up at the
first thermal pulse, thus the evolution along the TP--AGB phase is
not provided. Therefore, in the following we describe the
procedure presently adopted to simulate this bright and fast
evolutionary phase in our stellar population synthesis models in
order to properly compute SBF amplitudes.

Although it is well known that the AGB evolution of stellar masses
$m\lsim6$--$8\, M_{\sun}$ (depending on metallicity) ends with a
series of helium shell flashes or thermal pulses (TPs)
\citep[e.g.\ ][]{iben&renzini83}, a series of difficulties in
modelling these stars still arises. This depends from several
physical mechanisms. Among others, we recall: 1) the treatment of
convection, which is poorly known and commonly parameterized by
the Mixing Length Theory (MLT) with the ``mixing length scale''
free parameter (usually indicated as $\alpha$); 2) the mass--loss
rate, for which observations indicate a value ranging from
$10^{-7} M_{\sun}/yr$ to even $10^{-3} M_{\sun}/yr$ for the
coolest and luminous red super--giants \citep{Vanloon+99}; 3) the
occurrence of the Hot Bottom Burning (HBB), and of the third
dredge--up, which regulate the formation of carbon--rich stars
(C--stars); 4) the luminosity variations, which may reach an
amplitude of the order of few magnitudes in $V$ and few tenths in
NIR bands \citep{Cioni+01,Raimondo+05}. Therefore, fully modelling
the TP--AGB phase is indeed complex and time--consuming, thus
stellar population models make use of 'analytic' approaches
\citep{Wagenhuber&Groenewegen98,Marigo98}.

In intermediate and low mass stellar models by P04 the evolution
is followed from the MS until the onset of TPs. To extend the
track to the TP--phase we used the analytic formulas by WG98,
simulating the behavior of each TP--AGB star, in terms of its core
mass ($m_c$) and luminosity ($L$) as a function of time. The
adopted relation between the maximum bolometric luminosity $L$
during the quiescent hydrogen burning and the core mass $m_c$  (in
solar units) is:

\begin{eqnarray}
L \!=\! & &  (18160 + 3980 z)(\mc - 0.4468)\label{eqL}
 +  10^{2.705 \ + \ 1.649 \mc} \times \nonumber \\
&   & \times \left(10^{0.0237 (\alpha - 1.447) \mci2 \me2 (1 -
\e^{ - \Dmc / 0.01})}\right)
 -  10^{3.529 \-\ (m\I{c,0} - 0.4468) \Dmc / 0.01  }
\end{eqnarray}

\noindent where $z=log (Z/Z_{\sun})$ refers to the MS metal
abundance, $m_{c,0}$ is the core mass at the first thermal pulse,
$\Delta m_{c}$ is defined as $\Delta m_{c}= m_{c}-m_{c,0}$, and
$m_{e}$ is the envelope mass. Concerning the MLT parameter, we
assumed $\alpha=2$ according to the value used by P04 in the
previous evolutionary stages.

The equation of the core growth is:

\begin{equation}
\frac{\D m_c}{\D t} = \frac{q}{X\I{e}}L\I{H}, \label{eq:dmc}
\end{equation}
where $q$ is the mass burnt per unit energy release, and $X\I{e}$
is the hydrogen in the envelope (mass fraction):

\begin{equation}
q=[(1.02\pm0.04)+0.0017z]\times 10^{-11}
(M_{\sun}L_{\sun}^{-1}yr^{-1})
\end{equation}
$L\I{H}$ is the luminosity produced by H--burning, which is
obtained by the following equation:

\begin{equation}
log ~(L\I{H}/ L) = -0.012 - 10^{-1.25 - 113\Dmc} - 0.0016\me.
\end{equation}
Finally, the core mass--interpulse time relation is:
\begin{equation}
log~\tau\I{ip}  =  (-3.628 + 0.1337 z)(\mc - 1.9454) - 10^{-2.080
\-\ 0.353 z + 0.200 (\me + \alpha - 1.5)} - 10^{-0.626 \-\
70.30(\mci - z) \Dmc}
\end{equation}

We performed an integration of the system of equations obtaining,
for each thermal pulse, the luminosity, the core mass, and the
duration of the pulse. We also included a mass--loss rate
regulating the total mass at each TP star, as extensively discussed
in the next section. The stellar temperature is derived by using
prescription of Renzini \& Voli (1981), considering the
appropriate slope d$\ log (L/L_{\sun})$ /d$\ log(T_e)$ for the
evolutionary tracks we are using. The procedure ends by providing
the 'expected' evolution of a star of a given mass during the
TP--phase.

\subsection{Mass--loss efficiency for TP--AGB Stars}
\label{section:mass-loss}

In this work, mass loss is parameterized by following the prescription by
Reimers (1975) along the RGB and on AGB until the first thermal
pulse:
\begin{equation}
\dot m_{R} = - 4 \cdot 10^{-13}\eta_R \cdot LR/m
\label{eq:reimers}
\end{equation}
where $L, R, m$ are respectively the star luminosity, radius and
total mass in solar units; along these phases we assumed
$\eta_R=0.4$.

The quantitative determination of mass--loss rate along the
TP--phase is a complex problem, since dust formation and
circumstellar dust shells avoid in several case the possibility to
observe stars in the optical band. Nevertheless, the duration of
the TP--phase is triggered by the efficiency of the mass loss,
thus to parameterize mass--loss processes we followed several
prescriptions. After the first thermal pulse, we first used the
formulation by BH, derived from statistical properties of OH/IR
stars:

\begin{equation}
\dot m_{BH}= \mu LR/\me \label{eq:bh}
\end{equation}
with $\mu = -4\cdot 10^{-13}(m_{e,0}/m)$, being $ m_{e,0}$ the
envelope mass at the first TP. This is a modification of the
Reimers formula with $\eta_R =1$ which also includes a
dependence on the actual mass envelope.

For the sake of clearness, the present TP evolution is compared with the
results by other authors. In Table~\ref{tab:table_TP} we report
the TP evolution of a star with an original MS--mass $m_{MS}=7\,
M_{\sun}$, and solar metallicity with similar calculations by
Marigo (1998, M98), and Blocker \& Schoenberner (1991, BS91).
Table~\ref{tab:table_TP} lists thermal pulse number (Col.\ 1),
mass of the star (Col.\ 2), core mass (Col.\ 3), and the star
luminosity after a certain number of thermal pulses (Col.\ 4). The
first block refers to the TP evolution obtained by adopting the
procedure described above, starting from values of luminosity,
temperature, and core mass at the first TP according to P04 (case
$a$). The second block refers to the same TP procedure, but the
initial values of luminosity, temperature, and core mass at the
first TP are from BS91 (case $b$). The last two blocks report TP
evolution by BS91 and M98. There is a fair agreement (within $\sim
10 \%$ in luminosity) between block \#2, and blocks \#3 and \#4.
Larger differences are found if we make use of the stellar models
by P04 (case $a$). This case foresees stars experiencing the TP
stage at a level brighter than that found by M98 and BS91 (for
example $\sim 25 \%$ at TP \#30). Note that, for the given mass
($m_{MS}=7\, M_{\sun}$) tracks by P04 predict a higher luminosity
at the first TP than those adopted by M98 and BS91.

As shown by B95, the BH formulation cannot well reproduce
the observed initial--final mass relationship. He proposed a
mass--loss rate based on dynamical theoretical investigation on
the atmospheres of Mira--like variables by Bowen (1988). B95
suggested two mass--loss rates directly correlated with the
stellar luminosity:

\begin{equation}
\dot m_{B1} = 4.83 \cdot 10^{-9}m^{-2.1}_{MS}L^{2.7} \dot m_R
\label{eq:b1}
\end{equation}
\begin{equation}
\dot m_{B2} = 4.83 \cdot 10^{-9}m^{-2.1}_{TP}L^{2.7} \dot m_R
\label{eq:b2}
\end{equation}

\noindent with $\eta_R = 1$. These formulas are valid for long
period variable stars with periods $P>100$ d. They differ in the
adopted stellar mass: the case B1 makes use of the MS mass
($m_{MS}$), while B2 uses the actual mass ($m_{TP}$) predicting a
steeper increase of the mass--loss rate. As an example, in the
range of stellar masses we are interested in, the foreseen
mass--loss rates in the case of $m_{MS}=5M_\sun$, $Z=0.02$, at the
luminosity level $L\simeq 19600\ L_\sun$ are: $\dot m \simeq 6
\times 10^{-6}\ \Msun/yr$ in the BH case; $\dot m \simeq 4.4
\times 10^{-5}\ \Msun/yr$ for B1; and $\dot m \simeq 8.5 \times
10^{-5} \ \Msun/yr$ in the B2 case. The B95 scenarios foresee more
efficient mass--loss rates than BH, predicting a lower number of
stars and a shorter life--time along the TP--phase. The high
sensitivity of SBF to mass--loss is shown
Section~\ref{section:sbfandtp}, where we suggest that SBF might be used
as calibrator of mass--loss rate also for young and intermediate
stellar populations, similarly to old stellar systems (Paper I).

\section{About the stochastic effects on SBF measurements}
\label{section:appendix2}

\begin{figure*}
\epsscale{.80} \plotone{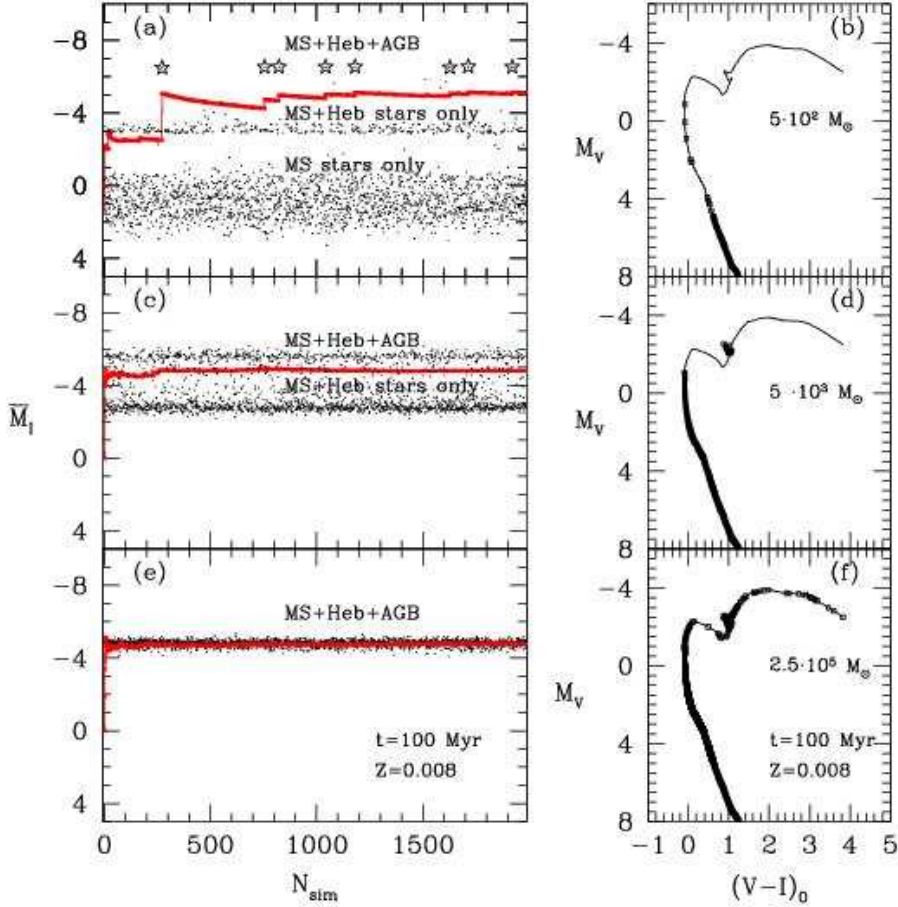}
\caption {Left panels: $\overline {M_I}$ for a population of
$100\, Myr$ and $Z=0.008$ as function of the simulation number
$N_{sim}$. The total mass of the stellar population increases from
the top to the bottom: ${\cal M}_{tot}=5 \cdot 10^2\ M_{\sun}$
(panel $a$); ${\cal M}_{tot}=5 \cdot 10^3\ M_{\sun}$ (panel $c$);
and ${\cal M}_{tot}=2.5 \cdot 10^5\ M_{\sun}$ (panel $e$). The
tick (red) solid lines refer to SBF computed from
\emph{std}--procedure. Differently, each dot represents the SBF as
derived from the $j$--th synthetic CMD ($M^{RS,j}_I$,
\emph{RS}--procedure). In panel $(a)$ the five--pointed stars at
$\overline {M}_I^{RS,j}\sim -6.5 \, mag$ enlighten the few
simulations in which also a few AGB stars are occasionally present
in the CMD. Right panels: the typical distribution of stars
(squares) in the synthetic CMD for the labelled total mass, and
the corresponding isochrone (solid line). Total masses are the
same as in the corresponding left panels. See the electronic
edition of the Journal for a color version of the figure. }
\label{fig:statistics2}
\end{figure*}

In order to deepen the matter of stochastic effects due to variations of
the number of bright stars, we performed numerical experiments by
varying the stellar population total mass $\cal M$.
Fig.~\ref{fig:statistics2}$(a,c,e)$ illustrates how $I$--SBF
amplitude is affected by stochastic effects depending on the star
cluster richness, as a function of $N_{sim}$. Solid (red) lines
refer to SBF derived by using the \emph{std}--procedure based on
integrated fluxes [Eq. \ref{eq:ourformula}: $\overline {M}_I^{std}
(N_{sim})$]; dots represent SBF derived from the
\emph{RS}--procedure to the $j-$th simulation (Eq.~\ref{eq:eqTS}:
$\overline {M}_I^{RS,j}$). We consider SSPs with $t=100\, Myr$ and
$Z=0.008$. The total mass increases from the top to the bottom.

First of all, we focus the attention on $\overline {M}_I^{RS,j}$. For
small values of the SSP mass, stars belong typically to the MS, as
indicated by the corresponding CMD
(Fig.~\ref{fig:statistics2}$b$). Due to the small number of stars,
only a few simulations have few occasional stars burning He in the
core, and even less simulations have the odd AGB stars.
Correspondingly, $\overline {M}_I^{RS,j}$ is dominated by MS stars
(Fig.~\ref{fig:statistics2}$a$, the bulk of dots at
${M}_I^{RS,j}\sim -1\, mag$). When few He--burning stars appear in
the CMD, the $I$--SBF becomes brighter ($\overline {M}_I^{RS,j}
\sim -3\, mag$), and even brighter if a TP--AGB star appears in
the simulation ($\overline {M}_I^{RS,j}\sim -6.5 \, mag$, the few
five--pointed stars).

By increasing the stellar population mass, the number of He--burning
stars and AGB stars grows in the CMD: first the He--burning phase
(Heb) becomes well populated (Fig.~\ref{fig:statistics2}$d$), then
for a further increase of the total mass a large number of stars
populates the AGB and TP--AGB phases
(Fig.~\ref{fig:statistics2}$f$). Correspondingly, the SBF signal
is always dominated by those evolved stars
(Fig.~\ref{fig:statistics2}$c,e$). Therefore, the scatter of the
SBF signal inversely correlates with the stellar population mass
$\cal M$: $\overline M_I^{RS,j}$ brighter than $-5\, mag$ are
never produced by SSPs with ${\cal M} = 2.5\cdot 10^5\,M_\odot $
while a few simulations among SSPs with ${\cal M} = 5\cdot 10^2\,
M_\odot$ can reach $\overline M_I \sim -6.5\, mag$. The reason is
that for lower $\cal M$ the fluctuation for the few simulations
with bright stars  is larger than the fluctuation of the typical
simulation without bright stars.

Red solid lines in Fig.~\ref{fig:statistics2} (left panels)
illustrate how the SBF signal derived using the
\emph{std}--procedure becomes stable  with $N_{sim}$ approaching
the \emph{asymptotic value}. Discontinuities in the solid lines
are directly related to the appearance of one or more very bright
stars in the simulation (e.g. Fig.~\ref{fig:statistics2}$a$). By
increasing the total mass of the population, this event occurs
with a higher probability, and discontinuities tend to disappear.
If we add stars to the population, we obtain a \emph{faster}
convergence of $\overline {M}_I^{RS,j}$ to the \emph{asymptotic
value} $\overline {M}_I^{std,asym}$
(Fig.~\ref{fig:statistics2}$c,e$).

Before closing this Appendix, let us make a further consideration
on Fig.~\ref{fig:statistics2}. From the figure, one can note that
$\overline {M}_I^{std}$ tends to converge rapidly to the
\emph{asymptotic value} already after $N_{sim} \sim 500$
simulations when ${\cal M}_{tot} \geq 5 \cdot 10^3\, M_{\sun}$.
However, even in the case of very poorly populated clusters
(Fig.~\ref{fig:statistics2}$a$) this happens before 2000
simulations. This feature has an important implication on the SBF
measurements of unresolved populations in distant galaxies. In
fact, the \emph{standard} procedure approaches the observational
way of deriving SBF in the case of unresolved stellar populations.
In such cases, the integrated flux we compute for each $j$--th
simulation corresponds to the flux measured in a single pixel of a
distant galaxy image (if one neglects seeing and population
mixture). Within the limit of this approximation, we find that
after cleaning processes (masking, galaxy subtraction, etc.) the
effective number of pixels used in deriving SBF measurements from
a galaxy CCD image should be larger than 2000, searching for the
presence of a young stellar population with a mass density lower
than $10^3\, M_{\sun}$/pixel. In case of higher density (say
$\gsim 10^4\, M_{\sun}/ $pixel), the constraint becomes less
severe and 500 pixels are enough to measure SBF with an
uncertainty lower than $0.1\, mag$.

\clearpage

\clearpage
\clearpage
\clearpage
\clearpage
\clearpage
\clearpage
\clearpage
\clearpage
\clearpage
\clearpage
\clearpage
\clearpage
\clearpage

\clearpage
\oddsidemargin=-1cm
\tabletypesize{\scriptsize}

\input{tab1.tex}

\clearpage

\input{tab2.tex}
\clearpage

\input{tab3.tex}
\clearpage

\input{tabA1.tex}

\end{document}

%% file: tab1.tex
\begin{deluxetable}{crrrrrrrrrrrrr}
\tablecaption{SBF amplitudes for $B1$--models} \tablewidth{380pt}
\tabletypesize{\tiny} \setlength{\tabcolsep}{0.05in}
\tablecolumns{9} \tablehead{ \colhead {Age (Gyr) }& \colhead {
$\bar {M}_U$ } & \colhead { $\bar {M}_B$ }& \colhead { $\bar
{M}_V$ }& \colhead { $\bar {M}_R$ } & \colhead { $\bar {M}_I$ }&
\colhead { $\bar {M}_J$ } & \colhead { $\bar {M}_H$ } & \colhead {
$\bar {M}_K$ } & \colhead { $\bar {M}_L$ } & \colhead {
$M_V^{Tot}$ } & \colhead { $V-I$ } } \startdata
\multicolumn {12} {c} {$Z=0.0003$ $ Y=0.245$}  \\
\tableline
   0.025&   $-$3.901&   $-$3.010&   $-$2.759&   $-$2.794&   $-$2.929&   $-$3.435&   $-$4.147&   $-$4.231&   $-$4.404&   $-$8.872&   -0.092  \\
   0.030&   $-$3.639&   $-$2.764&   $-$2.584&   $-$2.811&   $-$3.275&   $-$4.434&   $-$5.461&   $-$5.575&   $-$5.783&   $-$8.783&   -0.063  \\
   0.050&   $-$3.018&   $-$2.473&   $-$3.528&   $-$4.597&   $-$5.551&   $-$7.078&   $-$8.007&   $-$8.128&   $-$8.304&   $-$8.746&    0.276  \\
   0.080&   $-$2.441&   $-$2.041&   $-$3.013&   $-$4.051&   $-$5.006&   $-$6.555&   $-$7.547&   $-$7.675&   $-$7.863&   $-$8.572&    0.240  \\
   0.100&   $-$2.171&   $-$1.867&   $-$2.855&   $-$3.888&   $-$4.839&   $-$6.392&   $-$7.393&   $-$7.523&   $-$7.713&   $-$8.515&    0.252  \\
   0.200&   $-$1.450&   $-$1.326&   $-$2.303&   $-$3.284&   $-$4.192&   $-$5.711&   $-$6.718&   $-$6.845&   $-$7.042&   $-$8.312&    0.292  \\
   0.300&   $-$1.011&   $-$0.990&   $-$1.900&   $-$2.863&   $-$3.766&   $-$5.329&   $-$6.385&   $-$6.522&   $-$6.726&   $-$8.187&    0.320  \\
   0.400&   $-$0.645&   $-$0.761&   $-$1.503&   $-$2.377&   $-$3.232&   $-$4.759&   $-$5.822&   $-$5.959&   $-$6.170&   $-$8.121&    0.323  \\
   0.500&   $-$0.295&   $-$0.632&   $-$1.393&   $-$2.232&   $-$3.058&   $-$4.543&   $-$5.598&   $-$5.733&   $-$5.943&   $-$8.089&    0.351  \\
   0.600&   $-$0.098&   $-$0.458&   $-$1.300&   $-$2.142&   $-$2.948&   $-$4.387&   $-$5.414&   $-$5.550&   $-$5.751&   $-$8.109&    0.417  \\
   0.700&   $-$0.022&   $-$0.489&   $-$1.411&   $-$2.212&   $-$2.967&   $-$4.341&   $-$5.330&   $-$5.466&   $-$5.660&   $-$8.147&    0.539  \\
   0.800&      0.346&   $-$0.136&   $-$1.227&   $-$2.004&   $-$2.707&   $-$3.904&   $-$4.794&   $-$4.910&   $-$5.089&   $-$7.888&    0.595  \\
   0.900&      0.411&   $-$0.123&   $-$1.386&   $-$2.200&   $-$2.915&   $-$4.134&   $-$5.046&   $-$5.169&   $-$5.346&   $-$7.838&    0.662  \\
   1.000&      0.566&      0.050&   $-$1.217&   $-$2.004&   $-$2.689&   $-$3.825&   $-$4.682&   $-$4.795&   $-$4.965&   $-$7.755&    0.677  \\
   1.500&      0.843&      0.099&   $-$1.334&   $-$2.148&   $-$2.830&   $-$3.926&   $-$4.750&   $-$4.858&   $-$5.017&   $-$7.502&    0.761  \\
   2.000&      1.049&      0.343&   $-$1.061&   $-$1.866&   $-$2.544&   $-$3.646&   $-$4.478&   $-$4.588&   $-$4.749&   $-$7.232&    0.752  \\
   3.000&      1.137&      0.395&   $-$1.000&   $-$1.802&   $-$2.473&   $-$3.538&   $-$4.344&   $-$4.444&   $-$4.597&   $-$6.945&    0.774  \\
   4.000&      1.168&      0.334&   $-$1.077&   $-$1.882&   $-$2.555&   $-$3.641&   $-$4.452&   $-$4.554&   $-$4.708&   $-$6.764&    0.813  \\
   5.000&      1.206&      0.353&   $-$1.033&   $-$1.823&   $-$2.485&   $-$3.534&   $-$4.325&   $-$4.421&   $-$4.569&   $-$6.592&    0.833  \\
\tableline \multicolumn {12} {c} {$Z=0.001$ $ Y=0.246$}  \\
\tableline
   0.025&  $-$4.390&  $-$3.844&  $-$3.744&  $-$3.844&  $-$4.052&  $-$4.739&  $-$5.542&  $-$5.669&  $-$5.874&  $-$9.366&  -0.048  \\
   0.030&  $-$4.128&  $-$3.686&  $-$3.599&  $-$3.727&  $-$4.001&  $-$4.854&  $-$5.764&  $-$5.898&  $-$6.114&  $-$9.302&  -0.014  \\
   0.050&  $-$3.424&  $-$3.079&  $-$3.433&  $-$4.176&  $-$5.036&  $-$6.762&  $-$7.822&  $-$7.979&  $-$8.178&  $-$9.186&   0.267  \\
   0.080&  $-$2.713&  $-$2.484&  $-$2.876&  $-$3.599&  $-$4.436&  $-$6.154&  $-$7.271&  $-$7.438&  $-$7.653&  $-$8.987&   0.269  \\
   0.100&  $-$2.370&  $-$2.182&  $-$2.639&  $-$3.400&  $-$4.249&  $-$5.973&  $-$7.092&  $-$7.261&  $-$7.475&  $-$8.894&   0.299  \\
   0.200&  $-$1.386&  $-$1.469&  $-$2.014&  $-$2.815&  $-$3.672&  $-$5.429&  $-$6.574&  $-$6.747&  $-$6.966&  $-$8.574&   0.366  \\
   0.300&  $-$0.555&  $-$0.943&  $-$1.664&  $-$2.500&  $-$3.359&  $-$5.116&  $-$6.268&  $-$6.440&  $-$6.656&  $-$8.364&   0.410  \\
   0.400&  $-$0.010&  $-$0.486&  $-$1.320&  $-$2.131&  $-$2.959&  $-$4.694&  $-$5.863&  $-$6.039&  $-$6.259&  $-$8.191&   0.432  \\
   0.500&     0.319&  $-$0.081&  $-$1.062&  $-$1.896&  $-$2.720&  $-$4.437&  $-$5.603&  $-$5.776&  $-$5.997&  $-$8.090&   0.461  \\
   0.600&     0.546&     0.199&  $-$0.933&  $-$1.754&  $-$2.533&  $-$4.128&  $-$5.240&  $-$5.407&  $-$5.621&  $-$8.070&   0.543  \\
   0.700&     0.722&     0.192&  $-$1.172&  $-$2.002&  $-$2.742&  $-$4.188&  $-$5.226&  $-$5.388&  $-$5.583&  $-$8.140&   0.684  \\
   0.800&     1.057&     0.272&  $-$1.255&  $-$2.115&  $-$2.860&  $-$4.306&  $-$5.321&  $-$5.480&  $-$5.668&  $-$7.895&   0.732  \\
   0.900&     1.197&     0.367&  $-$1.170&  $-$2.026&  $-$2.763&  $-$4.157&  $-$5.147&  $-$5.302&  $-$5.487&  $-$7.780&   0.730  \\
   1.000&     1.276&     0.432&  $-$1.101&  $-$1.953&  $-$2.681&  $-$3.995&  $-$4.929&  $-$5.070&  $-$5.248&  $-$7.710&   0.734  \\
   1.500&     1.496&     0.525&  $-$1.038&  $-$1.897&  $-$2.625&  $-$3.941&  $-$4.865&  $-$5.004&  $-$5.181&  $-$7.475&   0.769  \\
   2.000&     1.674&     0.675&  $-$0.890&  $-$1.752&  $-$2.479&  $-$3.778&  $-$4.686&  $-$4.820&  $-$4.994&  $-$7.267&   0.785  \\
   3.000&     1.787&     0.731&  $-$0.813&  $-$1.672&  $-$2.399&  $-$3.746&  $-$4.674&  $-$4.812&  $-$4.989&  $-$6.925&   0.817  \\
   4.000&     1.876&     0.766&  $-$0.769&  $-$1.625&  $-$2.353&  $-$3.743&  $-$4.696&  $-$4.838&  $-$5.017&  $-$6.676&   0.845  \\
   5.000&     1.971&     0.861&  $-$0.655&  $-$1.505&  $-$2.230&  $-$3.634&  $-$4.594&  $-$4.736&  $-$4.916&  $-$6.472&   0.863  \\
\tableline \multicolumn {12} {c} {$Z=0.004$ $ Y=0.251$}  \\
\tableline
   0.025& $-$4.851& $-$4.708& $-$4.679& $-$4.840& $-$5.147& $-$5.962& $-$6.738& $-$6.865& $-$7.036& $-$10.077&  0.206  \\
   0.030& $-$4.219& $-$4.134& $-$4.193& $-$4.482& $-$4.931& $-$5.839& $-$6.566& $-$6.678& $-$6.829&  $-$9.854&  0.342  \\
   0.050& $-$3.323& $-$3.402& $-$3.550& $-$3.974& $-$4.659& $-$6.273& $-$7.415& $-$7.637& $-$7.875&  $-$9.562&  0.427  \\
   0.080& $-$2.071& $-$2.482& $-$2.817& $-$3.308& $-$4.052& $-$5.784& $-$6.999& $-$7.249& $-$7.509&  $-$9.238&  0.462  \\
   0.100& $-$1.397& $-$1.835& $-$2.412& $-$3.013& $-$3.821& $-$5.615& $-$6.850& $-$7.103& $-$7.366&  $-$9.059&  0.503  \\
   0.200&    0.007& $-$0.109& $-$1.234& $-$2.129& $-$3.150& $-$5.163& $-$6.459& $-$6.723& $-$6.990&  $-$8.525&  0.553  \\
   0.300&    0.487&    0.355& $-$0.802& $-$1.711& $-$2.746& $-$4.817& $-$6.161& $-$6.439& $-$6.718&  $-$8.287&  0.537  \\
   0.400&    1.070&    0.853& $-$0.546& $-$1.605& $-$2.731& $-$4.818& $-$6.132& $-$6.388& $-$6.647&  $-$8.070&  0.561  \\
   0.500&    1.436&    1.175& $-$0.254& $-$1.335& $-$2.473& $-$4.592& $-$5.916& $-$6.180& $-$6.444&  $-$7.918&  0.574  \\
   0.600&    1.708&    1.303& $-$0.202& $-$1.277& $-$2.344& $-$4.276& $-$5.518& $-$5.744& $-$5.980&  $-$7.835&  0.598  \\
   0.700&    1.807&    1.096& $-$0.510& $-$1.516& $-$2.471& $-$4.190& $-$5.336& $-$5.547& $-$5.769&  $-$7.895&  0.695  \\
   0.800&    1.852&    0.971& $-$0.685& $-$1.679& $-$2.607& $-$4.270& $-$5.379& $-$5.585& $-$5.801&  $-$7.952&  0.763  \\
   0.900&    2.027&    1.066& $-$0.640& $-$1.658& $-$2.616& $-$4.348& $-$5.476& $-$5.690& $-$5.918&  $-$7.665&  0.764  \\
   1.000&    2.120&    1.127& $-$0.575& $-$1.587& $-$2.523& $-$4.187& $-$5.277& $-$5.476& $-$5.690&  $-$7.569&  0.760  \\
   1.500&    2.477&    1.291& $-$0.431& $-$1.441& $-$2.373& $-$4.032& $-$5.115& $-$5.310& $-$5.521&  $-$7.273&  0.802  \\
   2.000&    2.615&    1.320& $-$0.375& $-$1.370& $-$2.315& $-$4.021& $-$5.125& $-$5.328& $-$5.545&  $-$7.122&  0.866  \\
   3.000&    2.789&    1.441& $-$0.232& $-$1.228& $-$2.194& $-$3.929& $-$5.041& $-$5.244& $-$5.458&  $-$6.775&  0.918  \\
   4.000&    2.900&    1.478& $-$0.184& $-$1.174& $-$2.145& $-$3.886& $-$4.997& $-$5.198& $-$5.411&  $-$6.497&  0.945  \\
   5.000&    2.977&    1.583& $-$0.052& $-$1.042& $-$2.037& $-$3.829& $-$4.958& $-$5.163& $-$5.379&  $-$6.279&  0.962  \\
\tableline \multicolumn {12} {c} {$Z=0.008$ $ Y=0.256$}  \\
\tableline
   0.025& $-$4.259& $-$4.523& $-$4.596& $-$4.841& $-$5.259& $-$6.121& $-$6.835& $-$6.963& $-$7.119& $-$10.214&  0.445 \\
   0.050& $-$1.598& $-$1.556& $-$2.816& $-$3.630& $-$4.521& $-$5.892& $-$6.959& $-$7.236& $-$7.548&  $-$9.402&  0.674 \\
   0.080& $-$0.855& $-$0.641& $-$1.906& $-$2.839& $-$3.994& $-$5.634& $-$6.840& $-$7.144& $-$7.483&  $-$9.058&  0.643 \\
   0.100& $-$0.561& $-$0.386& $-$1.597& $-$2.532& $-$3.756& $-$5.483& $-$6.739& $-$7.051& $-$7.398&  $-$8.912&  0.610 \\
   0.200&    0.126&    0.193& $-$0.757& $-$1.681& $-$3.083& $-$5.012& $-$6.384& $-$6.722& $-$7.092&  $-$8.497&  0.539 \\
   0.300&    0.843&    0.814& $-$0.262& $-$1.331& $-$2.903& $-$4.886& $-$6.273& $-$6.605& $-$6.966&  $-$8.221&  0.560 \\
   0.400&    1.332&    1.209&    0.110& $-$1.028& $-$2.658& $-$4.638& $-$6.027& $-$6.348& $-$6.701&  $-$8.008&  0.566 \\
   0.500&    1.680&    1.432&    0.206& $-$0.985& $-$2.543& $-$4.433& $-$5.767& $-$6.058& $-$6.381&  $-$7.873&  0.600 \\
   0.600&    1.962&    1.578&    0.203& $-$1.016& $-$2.548& $-$4.412& $-$5.728& $-$6.017& $-$6.338&  $-$7.769&  0.635 \\
   0.700&    2.147&    1.543&    0.018& $-$1.141& $-$2.543& $-$4.307& $-$5.569& $-$5.850& $-$6.164&  $-$7.773&  0.722 \\
   0.800&    2.273&    1.426& $-$0.225& $-$1.338& $-$2.528& $-$4.096& $-$5.218& $-$5.448& $-$5.700&  $-$7.806&  0.796 \\
   0.900&    2.377&    1.486& $-$0.178& $-$1.297& $-$2.572& $-$4.221& $-$5.407& $-$5.673& $-$5.978&  $-$7.522&  0.803 \\
   1.000&    2.491&    1.553& $-$0.127& $-$1.253& $-$2.530& $-$4.170& $-$5.345& $-$5.603& $-$5.901&  $-$7.443&  0.822 \\
   1.500&    2.903&    1.721&    0.016& $-$1.097& $-$2.411& $-$4.081& $-$5.276& $-$5.554& $-$5.870&  $-$7.113&  0.898 \\
   2.000&    3.142&    1.817&    0.089& $-$1.034& $-$2.358& $-$3.970& $-$5.132& $-$5.382& $-$5.667&  $-$6.895&  0.953 \\
   3.000&    3.357&    1.962&    0.276& $-$0.839& $-$2.232& $-$3.862& $-$5.043& $-$5.298& $-$5.594&  $-$6.573&  1.013 \\
   4.000&    3.503&    2.056&    0.398& $-$0.712& $-$2.181& $-$3.871& $-$5.085& $-$5.363& $-$5.683&  $-$6.317&  1.050 \\
   5.000&    3.659&    2.197&    0.554& $-$0.548& $-$2.048& $-$3.773& $-$5.010& $-$5.298& $-$5.630&  $-$6.080&  1.062 \\
\tableline \multicolumn {12} {c} {$Z=0.01$ $ Y=0.259$}  \\
\tableline
   0.025& $-$4.220& $-$3.903& $-$3.978& $-$4.386& $-$4.994& $-$6.064& $-$6.851& $-$6.991& $-$7.149& $-$9.914&  0.416 \\
   0.030& $-$3.775& $-$3.848& $-$3.987& $-$4.342& $-$4.877& $-$5.854& $-$6.618& $-$6.755& $-$6.911& $-$9.882&  0.464 \\
   0.050& $-$1.542& $-$1.176& $-$2.485& $-$3.417& $-$4.403& $-$5.884& $-$6.969& $-$7.283& $-$7.602& $-$9.346&  0.682 \\
   0.080& $-$0.910& $-$0.585& $-$1.755& $-$2.697& $-$3.884& $-$5.612& $-$6.823& $-$7.169& $-$7.515& $-$9.043&  0.621 \\
   0.100& $-$0.627& $-$0.419& $-$1.534& $-$2.454& $-$3.709& $-$5.505& $-$6.760& $-$7.109& $-$7.457& $-$8.920&  0.592 \\
   0.200&    0.191&    0.257& $-$0.680& $-$1.600& $-$3.040& $-$5.059& $-$6.432& $-$6.807& $-$7.177& $-$8.498&  0.531 \\
   0.300&    0.894&    0.909& $-$0.010& $-$1.050& $-$2.703& $-$4.809& $-$6.207& $-$6.577& $-$6.944& $-$8.189&  0.542 \\
   0.400&    1.365&    1.260&    0.297& $-$0.838& $-$2.594& $-$4.660& $-$6.045& $-$6.395& $-$6.747& $-$7.975&  0.558 \\
   0.500&    1.708&    1.446&    0.187& $-$1.073& $-$2.722& $-$4.618& $-$5.919& $-$6.232& $-$6.550& $-$7.843&  0.614 \\
   0.600&    2.023&    1.628&    0.261& $-$1.007& $-$2.658& $-$4.545& $-$5.843& $-$6.158& $-$6.478& $-$7.744&  0.653 \\
   0.700&    2.222&    1.567&    0.015& $-$1.154& $-$2.567& $-$4.277& $-$5.490& $-$5.782& $-$6.082& $-$7.765&  0.756 \\
   0.800&    2.347&    1.482& $-$0.167& $-$1.304& $-$2.624& $-$4.274& $-$5.448& $-$5.739& $-$6.039& $-$7.796&  0.846 \\
   0.900&    2.473&    1.572& $-$0.117& $-$1.292& $-$2.665& $-$4.314& $-$5.475& $-$5.754& $-$6.047& $-$7.511&  0.854 \\
   1.000&    2.590&    1.683& $-$0.002& $-$1.179& $-$2.566& $-$4.222& $-$5.390& $-$5.670& $-$5.964& $-$7.400&  0.859 \\
   1.500&    3.032&    1.877&    0.171& $-$0.974& $-$2.382& $-$4.085& $-$5.275& $-$5.578& $-$5.890& $-$7.037&  0.927 \\
   2.000&    3.317&    1.998&    0.273& $-$0.873& $-$2.319& $-$4.034& $-$5.226& $-$5.530& $-$5.843& $-$6.812&  0.982 \\
   3.000&    3.569&    2.132&    0.434& $-$0.698& $-$2.193& $-$3.943& $-$5.144& $-$5.452& $-$5.772& $-$6.504&  1.046 \\
   4.000&    3.688&    2.241&    0.580& $-$0.543& $-$2.072& $-$3.809& $-$5.013& $-$5.310& $-$5.623& $-$6.240&  1.077 \\
   5.000&    3.798&    2.284&    0.637& $-$0.474& $-$2.022& $-$3.826& $-$5.047& $-$5.366& $-$5.695& $-$6.035&  1.103 \\
\tableline
\multicolumn {12} {c} {$Z=0.0198$ $ Y=0.273$}  \\
\tableline
   0.025& $-$4.341& $-$4.161& $-$4.143& $-$4.356& $-$4.902& $-$5.894& $-$6.777& $-$6.961& $-$7.123& $-$10.042&  0.318 \\
   0.050& $-$2.683& $-$2.920& $-$3.080& $-$3.409& $-$4.134& $-$5.968& $-$7.145& $-$7.626& $-$7.970&  $-$9.545&  0.417 \\
   0.080& $-$1.435& $-$1.423& $-$2.031& $-$2.609& $-$3.571& $-$5.775& $-$7.021& $-$7.521& $-$7.871&  $-$9.172&  0.458 \\
   0.100& $-$0.887& $-$0.696& $-$1.442& $-$2.175& $-$3.291& $-$5.655& $-$6.925& $-$7.441& $-$7.800&  $-$8.979&  0.492 \\
   0.200&    0.237&    0.388& $-$0.171& $-$1.024& $-$2.530& $-$5.271& $-$6.587& $-$7.098& $-$7.456&  $-$8.433&  0.488 \\
   0.300&    0.937&    0.911&    0.291& $-$0.669& $-$2.396& $-$5.087& $-$6.387& $-$6.859& $-$7.192&  $-$8.134&  0.533 \\
   0.400&    1.418&    1.270&    0.500& $-$0.567& $-$2.432& $-$4.957& $-$6.232& $-$6.681& $-$7.000&  $-$7.915&  0.587 \\
   0.500&    1.826&    1.585&    0.692& $-$0.435& $-$2.326& $-$4.719& $-$5.983& $-$6.419& $-$6.731&  $-$7.754&  0.632 \\
   0.600&    2.136&    1.778&    0.680& $-$0.490& $-$2.334& $-$4.656& $-$5.899& $-$6.330& $-$6.639&  $-$7.644&  0.689 \\
   0.700&    2.435&    1.890&    0.487& $-$0.700& $-$2.309& $-$4.380& $-$5.564& $-$5.976& $-$6.273&  $-$7.628&  0.813 \\
   0.800&    2.619&    1.939&    0.444& $-$0.739& $-$2.333& $-$4.350& $-$5.519& $-$5.933& $-$6.230&  $-$7.594&  0.883 \\
   0.900&    2.739&    1.997&    0.422& $-$0.792& $-$2.432& $-$4.526& $-$5.673& $-$6.086& $-$6.379&  $-$7.358&  0.944 \\
   1.000&    2.869&    2.075&    0.453& $-$0.761& $-$2.381& $-$4.429& $-$5.575& $-$5.984& $-$6.276&  $-$7.225&  0.949 \\
   1.500&    3.382&    2.385&    0.702& $-$0.497& $-$2.131& $-$4.243& $-$5.391& $-$5.791& $-$6.076&  $-$6.817&  1.007 \\
   2.000&    3.770&    2.547&    0.838& $-$0.335& $-$1.930& $-$4.077& $-$5.231& $-$5.639& $-$5.927&  $-$6.574&  1.049 \\
   3.000&    4.109&    2.686&    0.971& $-$0.204& $-$1.849& $-$4.135& $-$5.289& $-$5.713& $-$6.010&  $-$6.256&  1.101 \\
   4.000&    4.250&    2.722&    1.049& $-$0.079& $-$1.651& $-$4.029& $-$5.187& $-$5.632& $-$5.939&  $-$6.024&  1.137 \\
   5.000&    4.396&    2.840&    1.177&    0.051& $-$1.548& $-$4.040& $-$5.192& $-$5.639& $-$5.943&  $-$5.817&  1.174 \\
\tableline \multicolumn {12} {c} {$Z=0.04$ $ Y=0.30$}  \\
\tableline
   0.025& $-$3.985& $-$4.192& $-$4.247& $-$4.393& $-$4.800& $-$5.750& $-$6.630& $-$6.870& $-$7.058& $-$10.056&  0.305 \\
   0.030& $-$3.390& $-$3.752& $-$3.931& $-$4.142& $-$4.664& $-$6.723& $-$7.857& $-$8.348& $-$8.671&  $-$9.936&  0.374 \\
   0.050& $-$1.901& $-$1.692& $-$2.450& $-$3.089& $-$3.875& $-$6.023& $-$7.214& $-$7.785& $-$8.173&  $-$9.413&  0.490 \\
   0.080& $-$1.166& $-$0.690& $-$1.206& $-$1.999& $-$3.076& $-$5.728& $-$6.952& $-$7.529& $-$7.915&  $-$8.996&  0.498 \\
   0.100& $-$0.828& $-$0.401& $-$0.796& $-$1.548& $-$2.698& $-$5.627& $-$6.890& $-$7.489& $-$7.884&  $-$8.831&  0.479 \\
   0.200&    0.259&    0.368&    0.099& $-$0.580& $-$1.934& $-$5.302& $-$6.621& $-$7.230& $-$7.628&  $-$8.329&  0.474 \\
   0.300&    0.954&    0.914&    0.560& $-$0.170& $-$1.602& $-$5.143& $-$6.456& $-$7.048& $-$7.431&  $-$8.015&  0.533 \\
   0.400&    1.549&    1.363&    0.840&    0.043& $-$1.590& $-$5.037& $-$6.322& $-$6.878& $-$7.243&  $-$7.785&  0.596 \\
   0.500&    1.972&    1.717&    1.030&    0.147& $-$1.616& $-$4.866& $-$6.127& $-$6.661& $-$7.016&  $-$7.607&  0.653 \\
   0.600&    2.319&    2.016&    1.147&    0.201& $-$1.576& $-$4.726& $-$5.974& $-$6.501& $-$6.850&  $-$7.461&  0.709 \\
   0.700&    2.594&    2.249&    1.156&    0.100& $-$1.801& $-$4.996& $-$6.188& $-$6.693& $-$7.026&  $-$7.328&  0.807 \\
   0.800&    2.727&    2.400&    1.188&    0.030& $-$1.607& $-$4.557& $-$5.768& $-$6.285& $-$6.625&  $-$6.919&  0.775 \\
   0.900&    2.915&    2.556&    1.232&    0.020& $-$1.698& $-$4.488& $-$5.674& $-$6.171& $-$6.498&  $-$6.811&  0.833 \\
   1.000&    3.104&    2.646&    1.225& $-$0.001& $-$1.744& $-$4.505& $-$5.676& $-$6.166& $-$6.492&  $-$6.718&  0.878 \\
   1.500&    3.695&    2.746&    1.158&    0.030& $-$1.572& $-$4.316& $-$5.463& $-$5.956& $-$6.282&  $-$6.585&  1.044 \\
   2.000&    4.063&    2.851&    1.232&    0.120& $-$1.467& $-$4.218& $-$5.358& $-$5.845& $-$6.163&  $-$6.334&  1.089 \\
   3.000&    4.466&    3.051&    1.422&    0.330& $-$1.224& $-$4.136& $-$5.286& $-$5.795& $-$6.123&  $-$5.992&  1.135 \\
   4.000&    4.662&    3.166&    1.546&    0.461& $-$1.094& $-$4.078& $-$5.230& $-$5.744& $-$6.077&  $-$5.761&  1.158 \\
   5.000&    4.796&    3.245&    1.633&    0.558& $-$0.979& $-$3.999& $-$5.152& $-$5.674& $-$6.012&  $-$5.590&  1.183 \\
\enddata \label{table:table_sbf}
\end{deluxetable}

%% file: tab2.tex
\begin{deluxetable}{lrcccllc}
\tablecolumns{5} \tablewidth{0pc} \tablecaption{Overview of the
WFPC2 observations for young clusters in the sample. }
\tabletypesize{\footnotesize} \tablehead{ \colhead { Cluster } &
\colhead {Archive Directory/File}  & \colhead { Filter }& \colhead
{ Total Exp. time (sec) } & \colhead { Date (DD/MM/YY)}}
\startdata
NGC 1805 & u4ax0204b    & F555W & 435 & 25/07/1998 \\
         & u4ax020ab    & F814W & 960 & 25/07/1998 \\
         & u4ax0501b    & F555W &7200 & 12/03/1998 \\
         & u4ax0601b    & F814W &4800 & 12/03/1998 \\
         & u4ax0803b    & F555W &2500 & 29/04/1998 \\
         & u4ax0903b    & F814W &2500 & 28/04/1998 \\
\hline
NGC 1818 & u4ax3004b    & F555W & 435 & 25/09/1998 \\
         & u4ax300ab    & F814W & 960 & 25/09/1998 \\
         & u4ax3301b    & F555W &7200 & 11/07/1998 \\
         & u4ax3501b    & F814W &4800 & 29/07/1998 \\
         & u4ax3603b    & F555W &2500 & 30/04/1998 \\
         & u4ax3703b    & F814W &2500 & 30/04/1998 \\
\hline
NGC 1868 & u4ax5204b    & F555W & 435 & 12/11/1998 \\
         & u4ax520ab    & F814W & 960 & 12/11/1998 \\
         & u4ax5301b    & F555W &7200 & 21/03/1998 \\
         & u4ax5601b    & F814W &4800 & 22/03/1998 \\
         & u4ax5803b    & F555W &2500 & 20/05/1998 \\
         & u4ax5903b    & F814W &2500 & 24/05/1998 \\
\hline
NGC 1831 & u4ax4104b    & F555W & 435 & 25/07/1998 \\
         & u4ax410ab    & F814W & 960 & 25/07/1998 \\
         & u4ax4401b    & F555W &7200 & 24/07/1998 \\
         & u4ax4601b    & F814W &4800 & 24/07/1998 \\
         & u4ax4703b    & F555W &2500 & 29/05/1998 \\
         & u4ax4803b    & F814W &2500 & 30/05/1998 \\
\hline
NGC 2209 & u4ax6304b    & F555W & 435 & 29/03/1998 \\
         & u4ax630ab    & F814W & 960 & 29/03/1998 \\
         & u4ax6401b    & F555W &7200 & 28/03/1998 \\
         & u4ax6701b    & F814W &4800 & 03/04/1998 \\
         & u4ax6903b    & F555W &2500 & 06/05/1998 \\
         & u4ax7003b    & F814W &2500 & 05/05/1998 \\
\hline
H14     & u4ax7404b    & F555W & 435 & 31/03/1998 \\
         & u4ax740ab    & F814W & 960 & 31/03/1998 \\
         & u4ax7301b    & F555W &1200 & 04/02/1998 \\
         & u4ax7303b    & F814W & 800 & 04/02/1998 \\
         & u4ax7501b    & F555W &7200 & 06/08/1998 \\
         & u4ax7801b    & F814W &4800 & 05/08/1998 \\
\enddata \label{tab:hstimages}
\end{deluxetable}

%% file: tab3.tex
\begin{deluxetable}{lrcccccc}
\tabletypesize{\footnotesize} \tablecolumns{9} \tablewidth{0pc}
\tablecaption{Clusters properties and measured SBF}
\tabletypesize{\small} \tablehead{ \colhead { $NGC$ } & \colhead {
$V_{tot}$\tablenotemark{a} } & \colhead { $V_{tot}$ } & \colhead {
$\bar {V}$ }  & \colhead { $\bar {I}$ }  & \colhead { $\bar {V} -
\bar {I}$ } & \colhead {log [t(yr)]} & \colhead {Ref.}} \startdata
1805 & 10.63  & 10.9   & 15.31$\pm$0.2 & 15.30$\pm$0.2 & 0.01$\pm$0.3 &   7.00$^{+0.30}_{-0.10}$ & 2  \\
1818 & 9.70   & 10.2   & 15.00$\pm$1.0 & 14.37$\pm$1.0 & 0.63$\pm$1.0 &   7.40$\pm$0.30          & 1  \\
1866 & 9.73   & 9.4    & 17.09$\pm$0.2 & 14.95$\pm$0.2 & 2.14$\pm$0.07&   8.15$\pm$ 0.30         & 3 \\
1831 & 11.18  & 10.9   & 18.47$\pm$0.2 & 15.90$\pm$0.2 & 2.57$\pm$0.3 &   8.65$\pm$ 0.30         & 1  \\
1868 & 11.56  & 11.5   & 18.68$\pm$0.5 & 16.50$\pm$0.5 & 2.18$\pm$0.5 &   8.81$\pm$ 0.30         & 1 \\
2209 & 13.15  & 12.5   & 18.75$\pm$0.5 & 16.66$\pm$0.5 & 2.09$\pm$0.5 &   8.85$\pm$ 0.20         & 1 \\
$Hodge$14&11.45& 13.7  & 19.21$\pm$1.0 & 17.29$\pm$1.0 & 1.92$\pm$0.5 &   9.30$\pm$ 0.10         & 1 \\
1754 & 11.86  & 12.2   & 18.42$\pm$0.5 & 16.70$\pm$0.5 & 1.72$\pm$0.4 &   10.19$^{+0.06}_{-0.07}$& 2  \\
1916 & 10.38  & 10.9   & 18.10$\pm$0.5 & 16.28$\pm$0.5 & 1.83$\pm$0.4 &   10.20$\pm$0.09         & 2  \\
2005 & 11.57  & 11.7   & 18.18$\pm$0.5 & 16.11$\pm$0.5 & 2.07$\pm$0.4 &   10.22$^{+0.12}_{-0.16}$& 2  \\
2019 & 10.86  & 11.4   & 17.94$\pm$0.5 & 16.13$\pm$0.5 & 1.81$\pm$0.4 &   10.25$^{+0.07}_{-0.09}$& 2 \\
\enddata \label{table:table_lmc}
\tablecomments{References for the cluster age: 1=present work;
2=MG03; 3=Brocato et al. (2003)} \tablenotetext{a}{Van den Bergh
(1981)}
\end{deluxetable}

%% file: tabA1.tex
\begin{deluxetable}{rrrr}
\tablenum{A.1} \tablecolumns{4} \tablewidth{0pc}
\tablecaption{Comparison with previous works. $BH$ mass--loss rate
is adopted.} \tabletypesize{\footnotesize} \tablehead{ \colhead {
$No.$ } & \colhead { $m/M_{\sun}$ }   & \colhead { $m_{c}/M_{\sun}
$} & \colhead { $L/L_{\sun}$ }   } \startdata \multicolumn {4} {c}
{Present procedure (case $a$)} \\ \tableline
1     &  6.928 &  0.938   & 33798 \\
10    &  6.903 &  0.944   & 48265 \\
20    &  6.837 &  0.953   & 64887 \\
30    &  6.674 &  0.961   & 73937 \\
\tableline \multicolumn {4} {c}{Present procedure (case $b$)}\\
\tableline
1     &  6.871 &  0.913   & 25217 \\
10    &  6.848 &  0.920   & 36581 \\
20    &  6.796 &  0.930   & 48570 \\
30    &  6.728 &  0.940   & 54352 \\
\tableline \multicolumn {4} {c}{BS91 }\\ \tableline
1     &  6.871 &  0.913  & 25217 \\
10    &  6.854 &  0.921  & 40268 \\
20    &  6.821 &  0.930  & 51662 \\
30    &  6.774 &  0.939  & 58806 \\
\tableline \multicolumn {4} {c}{Marigo (1998)}\\\tableline
1     &  6.871 &  0.913  & 25217 \\
11    &  6.850 &  0.921  & 37788 \\
22    &  6.807 &  0.930  & 51537 \\
34    &  6.745 &  0.939  & 60760 \\
\enddata
\label{tab:table_TP}
\end{deluxetable}

%% file: ms.bbl
\clearpage

\begin{thebibliography}{}

\bibitem[Alcock et al. 2004]{Alcock+04} Alcock, C., Alves, D.\ R., Axelrod, T.\ S., Becker, A.\ C., Bennett,
D.\ P., Clement, C.\ M., Cook, K.\ H., Drake, A.\ J., Freeman, K.\
C., et al.  2004, AJ, 127, 334

\bibitem[Aaronson \& Mould 1982]{Aaronson&Mould82} Aaronson, M., \&
    Mould, J. 1982, ApJSS, 48, 161

\bibitem[Ajhar \& Tonry 1994]{Ajhar&Tonry94} Ajhar, E.\ A., \& Tonry, J.\ L.
     1994, ApJ, 429, 557 (AT94)

\bibitem[Angulo et al. 1999]{Angulo+99} Angulo, C., Arnould, M.,
Rayet, M., et al. 1999 (NACRE collaboration) Nucl. Phys. A, 656, 3

\bibitem[Baud \& Habing 1983]{baud&habing83} Baud, B., \& Habing, H.\ J. 1983,
    \aap, 127, 73 (BH)

\bibitem[Blakeslee, Vazdekis, and Ajhar 2001]{blakeslee+01} Blakeslee, J.\ P.,
    Vazdekis, A., Ajhar, E.\ A. 2001, \mnras, 320, 193

\bibitem[Blocker \& Schoenberner 1991]{Blocker&schoenberner91}
    Blocker, T., \& Schoenberner, D. 1991, \aap, 244, L43 (BS1)

\bibitem[Blocker 1995]{Blocker95} Blocker, T.  1995, A\&A, 297,
    727 (B95)

\bibitem[Bowen 1988]{Bowen88} Bowen , G.H. 1988, ApJ, 329, 299

\bibitem[Brocato et al. 1998]{Brocato+98} Brocato, E., Capaccioli, M., Condelli,
M. 1998, Mem. SAIt, Vol. 69, p.155

\bibitem[Brocato et al. 2003]{Brocato+03} Brocato, E., Castellani, V.,
    Di Carlo, E., Raimondo, G., Walker, A. R. 2003, AJ, 125, 3111

\bibitem[Brocato et al. 2000]{Brocato+00} Brocato, E., Castellani, V.,
    Poli, F.M.,  \& Raimondo, G. 2000, A\&AS, 146, 91

\bibitem[Brocato et al. 1999]{Brocato+99} Brocato, E., Castellani, V.,
    Raimondo,  \& G., Romaniello M. 1999, A\&AS, 136, 65

\bibitem[Buzzoni 1993]{buzzoni93} Buzzoni, A. 1993, \aap, 275, 433

\bibitem[Cantiello et al. 2003]{Cantiello+03} Cantiello, M., Raimondo, G.,
    Brocato, E., \& Capaccioli, M. 2003, \aj, 125, 2783 (Paper I)

\bibitem[Cantiello et al. 2005]{Cantiello+05} Cantiello, M., Blakeslee, J. P., Raimondo, G.,
Mei, S., Brocato, E.,  Capaccioli, M. 2005, ApJ in press, astro-ph

\bibitem[Carpenter 2001]{Carpenter01} Carpenter, J.\ M. 2001, AJ, 121, 2851

\bibitem[Castellani et al. 2003]{Castellani+03} Castellani, V.,
 Degl'Innocenti, S., Marconi, M., Prada Moroni, P. G., Sestito, P.
 2003, A\&A, 404, 645

\bibitem[Cioni et al. 2003]{Cioni+03} Cioni, M.-R. L.,
Blommaert, J. A. D. L., Groenewegen, M. A. T., Habing, H. J.,
Hron, J., Kerschbaum, F., Loup, C., Omont, A., van Loon, J. Th.,
Whitelock, P. A., Zijlstra, A. A 2003, A\&A, 406, 51

\bibitem[Cioni et al. 2001]{Cioni+01} Cioni, M.-R.\ L., Marquette, J.-B.,
    Loup, C., Azzopardi, M., Habing, H.\ J., Lasserre, T., Lesquoy, E.
    2001, A\&A, 377, 945

\bibitem[Cohen  1982]{Cohen+85} Cohen J.\ G. 1982 ApJ, 258, 143


\bibitem[de Grijs et al. 2002a]{deGrijs+02a} de Grijs, R., Johnson, R.\ A., Gilmore,
    G.\ F., Frayn, C.\ M. 2002a, MNRAS, 331, 228

\bibitem[de Grijs et al. 2002b]{deGrijs+02b} de Grijs, R., Gilmore, G. F., Johnson, R. A., Mackey, A. D.
2002b, MNRAS ,331, 245

\bibitem[de Grijs et al. 2002c]{deGrijs+02c} de Grijs, R., Gilmore, G. F., Mackey, A. D., Wilkinson, M. I.,
Beaulieu, S. F., Johnson, R. A., Santiago, B. X. 2002c, MNRAS,
337, 597

\bibitem[Dolphin 2000a]{Dolphin00a}Dolphin, A. E. 2000a, PASP, 112, 1383

\bibitem[Dolphin 2000b]{Dolphin00b} Dolphin, A. E. 2000b, PASP, 112, 1397

\bibitem[Elson et al. 1985]{Elson+85} Elson, R.\ A.\ W. \& Fall
S.\ M. 1985, ApJ, 299, 211

\bibitem[Elson et al. 1988]{Elson+88} Elson, R.\ A.\ W. \& Fall
S.\ M. 1988, AJ, 96, 1383

\bibitem[Frogel and Cohen 1982]{Frogel+82} Frogel J.\ A., Cohen
J.\ G. 1982, ApJ, 253, 580

\bibitem[Girardi et al. 1995]{Girardi+95}Girardi, L., Chiosi, C., Bertelli, G., Bressan, A.
1995, A\&A, 298, 87

\bibitem[Girardi et al. 2000]{Girardi+00} Girardi, L., Bressan, A., Bertelli, G., Chiosi, C.
2000, A\&AS 141, 371

\bibitem[Gonzalez et al. 2003]{Gonzalez+03} Gonzalez, R.\ A., Liu, M.\ C.,
    Bruzual, G.\ A., 2003, poster paper in Stellar
    Populations 2003, Conf. Garching, Germany, Oct. 6--10, 2003

\bibitem[Gonzalez et al. 2004]{Gonzalez+04} Gonzalez, R.\ A., Liu, M.\ C.,
    Bruzual, G.\ A. 2004,  \apj, 611, 27 (G04)

\bibitem[Gonzalez et al. 2005]{Gonzalez+05} Gonzalez, R.\ A., Liu, M.\ C.,
    Bruzual, G.\ A. 2005,  \apj, 621, 557

\bibitem[Groenewegen et al. 1995]{Groenewegen+95} Groenewegen, M.\ A.\ T., Smith, C.\ H., Wood, P.\ R., Omont, A.,
Fujiyoshi, T.  1995, \apj, 449, L119


\bibitem[Herwig et al. 1997]{Herwig+97} Herwig, F., Blocker, T., Schoenberner, D., El Eid, M.
    1997, A\&A, 324, L81

\bibitem[Kunz et al. 2002]{Kunz+02}Kunz, R., Fey, M.,
Jaeger, M., Mayer, A., Hammer, J. W., Staudt, G., Harissopulos,
S., Paradellis, T. 2002, ApJ, 567, 643

\bibitem[Iben \& Renzini 1983]{iben&renzini83} Iben, I., Jr., \&
    Renzini, A. 1983,
    \araa, 21, 271

\bibitem[Irwin et al. 2004]{Irwin+04} Irwin A.\, W., et al. 2004 in
preparation

\bibitem[Jensen et al. 2003]{jensen+03} Jensen J.B., Tonry, J.L.,
    Barris, B.J., Thompson, R.I., Liu, M.C., Rieke, M.J.,
    Ajhar, E.A., Blakeslee, J.P., 2003, \apj, 583, 712


\bibitem[Lejeune et al. 1997]{lejeune+97} Lejeune, T., Cuisinier, F.,
    \& Buser, R. 1997, A\&AS, 125, 229

\bibitem[Liu, Charlot, and Graham 2000]{Liu+00} Liu, M.C., Charlot, S., \&
    Graham, J.R. 2000, \apj, 543, 644

\bibitem[Liu, Graham, \& Charlot  2002]{Liu+02} Liu, M.C., Graham, J.R.,  \&
    Charlot S. 2002, \apj, 564, 216


\bibitem[MG03]{Mackey&Gilmore03} Mackey, A.\ D., \& Gilmore, G.\ F.
     2003, MNRAS, 338, 85 (MG03)

\bibitem[Marigo 1998]{Marigo98} Marigo, P. 1998, A\&A, 340, 463
(M98)

\bibitem[Marigo et al. 2003]{Marigo+03} Marigo, P., Girardi, L., Chiosi, C.
 2003, A\&A, 403, 225

\bibitem[Mei, Quinn \& Silva  2001]{mei+01} Mei, S., Quinn, P. J., \&
    Silva, D. R. 2001, \aap, 371, 979


\bibitem[Nikolaev \& Weinberg 2000]{Nikolaev&Weinberg00} Nikolaev, S., \& Weinberg, M.
D.  2000, \apj, 542, 804

\bibitem[Olsen et al. 1998]{Olsen+98} Olsen, K.\ A.\ G., Hodge, P.\ W., Mateo,
M., Olszewski, E.\ W., Schommer, R.\ A., Suntzeff, N.\ B., Walker,
A.\ R. 1998, MNRAS, 300, 665


\bibitem[Pahre at al. 1999]{Pahre+99} Pahre, M.\ A., Mould, J.\ R., Dressler, A.\, Holtzman,
J.\ A., Watson, A.\ M., Gallagher, J.\ S., III; Ballester, G. E.,
Burrows, C.\ J., et al.  1999, \apj, 515, 79

\bibitem[Pagel \& Edmunds 1981]{Pagel&Edmunds} Pagel, B.\ E.\ J., \& Edmunds, M.\
G. 1981, ARA\&A, 19, 77

\bibitem[Persson et al. 1983]{Persson+83}
Persson, S.\ E., Aaronson, M., Cohen, J.\ G., Frogel, J.\ A.,
Matthews, K., 1983, ApJ, 266, 105

\bibitem[Pietrinferni et al.  2004]{Pietrinferni+04} Pietrinferni, A.,
    Cassisi, S., Salaris, M., Castelli, F. 2004, \apj,  612, 168
    (P04)

\bibitem[Pols at al. 2001]{Pols+01} Pols, O.\ R., \& Tout, C.\ A. 2001, MmSAI, 72, 299

\bibitem[Raimondo et al. 2003]{Raimondo+03} Raimondo, G.,
    Brocato, E., Cantiello, M., \& Capaccioli, M. 2003,
    poster paper in Stellar
    Populations 2003, Conf. Garching, Germany, Oct. 6--10, 2003

\bibitem[Raimondo et al. 2004]{Raimondo+04} Raimondo, G., Cantiello, M.,
Brocato, E., Capaccioli, M. 2004, MmSAIt, 75, 198

\bibitem[Raimondo et al. 2005]{Raimondo+05} Raimondo, G., Cioni,
M.\ R.\ L., Rejkuba, M., \& Silva, D.\ R. 2005, A\&A in press,
astro-ph 0503561

\bibitem[Reimers 1975]{Reimers75} Reimers, D. 1975, in Problems in stellar atmospheres and envelopes,
 New York, Springer-Verlag, p. 229--256.

\bibitem[Renzini \& Buzzoni 1986]{Renzini&Buzzoni86} Renzini, A., \& Buzzoni, A. 1986,
    in Proceedings of the Fourth Workshop, Erice,
    Italy, March 12-22, 1985, Spectral evolution of galaxies
    (Dordrecht, D. Reidel Publishing Co.),195

\bibitem[Renzini \& Voli  1981]{Renzini&Voli81} Renzini, A. \& Voli, M. 1981,
    \aap, 94, 175

\bibitem[Salaris et al. 2003]{Salaris+03} Salaris, M., Percival, S., Brocato, E.,
    Raimondo, G., \& Walker, A. R. 2003, ApJ, 588, 801

\bibitem[Santos \& Frogel 1997]{Santos&Frogel97} Santos, J.\ F.\ C.\ Jr., \& Frogel, J.\ A.
    1997, ApJ, 479, 764

\bibitem[Scalo 1998]{Scalo98} Scalo, J.M. 1998, The Stellar Initial Mass Function
(38th Herstmonceux Conference) ed. by G. Gilmore and D. Howell,
ASP Conf. Ser., Vol. 142, 1998, p.201

\bibitem[Searle et al. 1980]{Searle+80} Searle L., Wilkinson A.,
Bagnuolo W.\ G. 1980 ApJ, 239, 803

\bibitem[Sodemann \& Thomsen 1995]{Sodemann&Thomsen95} Sodemann, M., \& Thomsen, B.
    1995, AJ, 110, 179

\bibitem[Stephens et al. 2001]{Stephens+01} Stephens, A.W., Frogel,
J.A., Freedman, W., Gallart, C., Jablonka, P., Ortolani, S.,
Renzini, A., Rich, R.M., Davies, R.M. 2001, AJ, 121, 2584

\bibitem[Straniero et al. 1997]{Straniero+97} Straniero, O.,
    Chieffi, A., Limongi, M., Busso, M., Gallino, R., Arlandini, C.
    1997, \apj, 478, 332

\bibitem[Tonry et al. 2001]{tonry+01} Tonry, J.L., Dressler, A.,
    Blakeslee, J.P.,
     Ajhar, E. A., Fletcher, A. B., Luppino, G.A., Metzger, M.R.,
    \& Moore, C. B. 2001, \apj, 546, 681

\bibitem[Tonry \& Schneider 1988]{ts88} Tonry, J.L., \& Schneider D.P.
    1988, \aj, 96, 807

\bibitem[van den Bergh 1981]{vandenBergh81} van den Bergh, S.  1981, A\&AS, 46, 79

\bibitem[Van Loon et al. 1999]{Vanloon+99} Van Loon, J.\ Th.,
    Groenewegen, M.\ A.\ T., de Koter, A., Trams, N.\ R.,
    Waters, L.\ B.\ F.\ M., Zijlstra, A.\ A., Whitelock, P.\ A., Loup, C.
    1999, A\&A, 351, 559

\bibitem[WG98]{Wagenhuber&Groenewegen98}
    Wagenhuber, J., \& Groenewegen, M.\ A.\ T.\ 1998, A\&A, 340,
    183 (WG98)

\bibitem[Walker et al. 2001]{Walker+01} Walker, A.\ R., Raimondo, G., Di\ Carlo, E., Brocato,
E., Castellani, V., \& Hill, V. 2001, ApJ, 560, L139

\bibitem[Westera et al. 2002]{Westera+02} Westera, P., Lejeune, T., Buser, R.,
    Cuisinier, F., \& Bruzula, G. 2002, A\&A, 381, 524

\bibitem[Westerlund  1997]{Westerlund97} Westerlund, B.\ E.
1997: \emph{The Magellanic Clouds}, Cambridge University Press

\bibitem[Westerlund et al. 1991]{Westerlund+91} Westerlund, B.\ E., Azzopardi, M.,
    Rebeirot, E., Breysacher, J.  1991, A\&AS, 91, 425

\bibitem[Worthey 1993]{Worthey93} Worthey, G. 1993, \apj, 415, 91

\end{thebibliography}
